\documentclass[aps,prd,twocolumn,superscriptaddress]{revtex4-2}

\usepackage{makecell} 
\usepackage{microtype}
\usepackage{graphicx}
\usepackage{dcolumn}
\usepackage{bm}
\usepackage{caption} 
\usepackage{mathtools} 
\usepackage{tensor} 
\usepackage[hidelinks]{hyperref} 
\usepackage{amssymb}
\usepackage{mathtools}
\usepackage{mathrsfs}
\usepackage{amsmath} 
\usepackage{xcolor}
\usepackage{array} 
\usepackage{booktabs}
\usepackage{subcaption}
\usepackage{enumitem}
\usepackage{soul}
\usepackage{amsthm}
\usepackage{stmaryrd}

\usepackage{amsfonts} 

\newcommand{\Ree}{\mathfrak{R}}
\newcommand{\Imm}{\mathfrak{I}}

\usepackage[toc,page]{appendix}

\theoremstyle{remark}

\newcommand{\dd}{\mathrm{d}}

\newcommand{\s}{\mathfrak{s}}

\newcommand{\schdual}{\widetilde{\mathfrak{s}}}
\newcommand{\al}{\alpha_0}

\begin{document}

\title{Harmonic branches of Einstein--Maxwell--scalar field gravity: exact solutions and regularity}

\author{Leonel Bixano}
    \email{Contact author: leonel.delacruz@cinvestav.mx}
\author{Tonatiuh Matos}%
 \email{Contact author: tonatiuh.matos@cinvestav.mx}
\affiliation{Departamento de F\'{\i}sica, Centro de Investigaci\'on y de Estudios Avanzados del Instituto Politécnico Nacional, Av. Instituto Politécnico Nacional 2508, San Pedro Zacatenco, M\'exico 07360, CDMX.
}%

\date{\today}

\begin{abstract}
We relate black-hole horizons and wormhole throats to the geometry of the harmonic branches of Einstein--Maxwell--dilaton gravity. In the unified Ernst framework, we first solve the harmonic equations prior to fixing the spatial harmonic, and then jointly reconstruct the metric, Maxwell four-potential, and scalar field. The conserved geodesic energy fixes the meridional Weyl quadrature, whereas the initial geodesic data capture the asymptotic charges. Choosing the Schwarzschild harmonic function, regular nonextremal horizons fix the normalization \(H=1\) along with the associated matter-field regularity requirements. This reproduces the Schwarzschild and Reissner-Nordström families (including the NUT extension), the Gibbons-Maeda solution, and the dyonic string and Kaluza--Klein solutions, while also preserving their more general branches. If the scalar sector is continued to the phantom sign, one obtains a complementary oblate construction: smooth continuation selects \(H=-1\) and produces the massive and ultrastatic Ellis–Bronnikov wormholes. By comparing with the vacuum ring geometry, we make clear the further importance of the global extension. Altogether, these findings offer a unified blackhole-wormhole geometric reading of the charge relations and regularity constraints for the reconstructed solutions.
\end{abstract}

\maketitle

\section{Introduction}
The Ernst potentials reformulate the stationary, axisymmetric Einstein equations into a framework whose complex structure makes clear connections among exact solutions that might otherwise appear unrelated \cite{Ernst:1967by,Ernst:1967wx}. In Einstein--Maxwell--dilaton (EMD) gravity, the harmonic potential method can be separated on to two steps: The initial step is to determine the affine parametrized potential-space geodesic associated with a given branch, then select the matching harmonic function that encodes the gravitational multipoles, thereby constructing the known solutions. Following this strategy, Refs \cite{Matos:1995my,Matos:2000za,Bixano:2026ouq,Bixano:2026xum} constructed static classes based on one or two harmonics, as well as stationary dilatonic setups. Although additional potential-space symmetries arise for particular choices of the coupling, one can analyze the harmonic field equations without restricting to those special values \cite{Bixano:2026duf}.

We begin by introducing the field equations obtained in \cite{Bixano:2026xum}.Except in Sec. \ref{sec:wormholes}, we work within the dilatonic scalar sector, \(\epsilon_0=+1\), using the Einstein-frame action
\begin{equation}
     S=\frac1{16\pi}\int\dd^4x\sqrt{-g}
     \left[-R+2(\nabla\phi)^2+e^{-2\al\phi}F_{\mu\nu}F^{\mu\nu}\right],
     \label{eq:action}
\end{equation}
with \(F=\dd\mathcal A\), \(G=c=1\), and the metric and four-potential chosen as
\begin{equation}
     \begin{aligned}
     \dd s^2={}&-f(\dd t-\omega\dd\varphi)^2
     +f^{-1}\left[e^{2k}(\dd\rho^2+\dd z^2)+\rho^2\dd\varphi^2\right],\\
     \mathcal A={}&A_t\dd t+A_\varphi\dd\varphi.
     \end{aligned}
     \label{eq:weyl}
\end{equation}
Here all metric functions and potentials are functions of \(\rho,z\). We adopt the conventions \(\psi=2A_t\), \(\kappa=e^{-\al\phi}\), together with the twist and magnetic potentials \(\epsilon,\chi\) as defined in Ref. \cite{Bixano:2026xum}. As a second input we use the Hamiltonian and Noether reconstruction developed in \cite{Bixano:2026duf}, whose Maxwell sector provides the quadratures employed later. Accordingly, we take the underlying formalism as given, while carrying out the integrations and the physical interpretation of the harmonic branches in the present work.

We use these conserved quantities to connect the geometry in potential space with regularity at horizons and throats. For each branch, we keep the integration constants explicit until imposing the boundary conditions. This makes it clear which charge relations follow from the geodesic energy and which are instead required by matter regularity. The oblate sector provides an additional comparison through the massive and ultrastatic Ellis--Bronnikov wormholes \cite{Ellis:1973yv,Bronnikov:1973fh}. We further compare their smooth continuation to the vacuum ring geometry \cite{Gibbons:2016bok}, whose global completion necessitates a distributional source.

\section{Potentials method}\label{Sec:PotentialMethod}

\subsection{Harmonic field equations}\label{sec:reconstruction}

On the auxiliary flat subspace defined by \(\sigma=0\), one may select an arbitrary real Weyl-harmonic function \(\s(\rho,z)\), i.e. satisfying \(\rho^{-1}D\!\cdot(\rho D\s)=0\) with \(D=(\partial_\rho,\partial_z)\). Assuming that all potentials are functions of \(\s\), the one-forms take the form \(A=a(\s)D\s\), \(B=b(\s)D\s\), and \(C=c(\s)D\s\). In this case the three field equations reduce to (see \cite{Bixano:2026xum})
\begin{equation}
     \begin{aligned}
     a'&=b\bar b+a(a-\bar a),\\
     b'&=-\tfrac12b(a-3\bar a)+c\bar b,\\
     c'&=\tfrac{\al^2}{2}(b^2+\bar b^{\,2}).
     \end{aligned}
     \label{eq:abc}
\end{equation}
where a prime indicates differentiation with respect to \(\s\), \(\dd/\dd\s\). The coefficients connect to the potentials through \(a=[f'-i(\epsilon'-\psi\chi')]/(2f)\), \(b=-(\kappa\psi'-i\kappa^{-1}\chi')/(2\sqrt f)\), and \(c=-(\ln\kappa)'\). Hence, the corresponding reconstruction reads
\begin{equation}
     \begin{aligned}
     (\ln f)'&=2\Ree a,&(\ln\kappa)'&=-c,\\
     A_t'&=-\frac{\sqrt f}{\kappa}\Ree b,
     &\chi'&=2\kappa\sqrt f\,\Imm b,\\
     \epsilon'&=\psi\chi'-2f\Imm a,
     &\phi'&=\frac{c}{\al}\quad(\al\ne0).
     \end{aligned}
     \label{eq:reconstruct}
\end{equation}
More specifically, \(\Ree a\) (Real part of the variable \(a\)) sets the redshift while \(\Imm a\) (Imaginary part of the variable \(a\)) sets the twist. The real and imaginary components of \(b\) fix the electric and magnetic potentials, and \(c\) specifies the scalar field. For \(\al=0\), the scalar field is uncoupled from the electromagnetic field. One may still allow an independent scalar field to remain nonzero or, when appropriate, set it to zero, i.e., \(C=0\). In both pure electrovacuum and vacuum, one always assumes \(\phi=\phi_\infty=0\).

For a harmonic function with vanishing asymptotic behavior, i.e. \(\lim_{x\rightarrow \infty} \s=0\), we can adopt the normalization \(f(0)=\kappa(0)=1\) and \(A_t(0)=\chi(0)=0\), and fix \(\s=0\) at the selected point at infinity. Here \(x\) is related to the radial Boyer–Lindquist coordinate, which is introduced in the next subsection \ref{sec:coordinates}. Imposing this condition on \(\kappa\) determines the asymptotic dilaton vacuum for the charged EMD families. Alternative vacua can be recovered via the shifts \(\phi\mapsto\phi+\phi_\infty\) and \(\mathcal A\mapsto e^{\al\phi_\infty}\mathcal A\). The corresponding flux relations imply \(q_0\mapsto e^{-\al\phi_\infty}q_0\) and \(p_0\mapsto e^{\al\phi_\infty}p_0\). In this way the normalization is fixed while retaining independent parameters for the mass, NUT charge, and electromagnetic charge.

For \(\al\neq 0\), Eq \eqref{eq:abc} admits the conserved quantity
\begin{equation}
     H=|a|^2-|b|^2+\frac{c^2}{\al^2}.
     \label{eq:energy}
\end{equation}
In the purely electrovacuum situation, \eqref{eq:energy} reduces to \(H=|a|^2-|b|^2\). To verify its conservation, we differentiate the three terms separately. Equation \eqref{eq:abc} gives \((|a|^2)'=2\Ree(\bar a a')=2(\Ree a)|b|^2\), \((|b|^2)'=2\Ree(\bar b b')=2(\Ree a)|b|^2+2c\Ree(b^2)\), and \((c^2/\al^2)'=2c\Ree(b^2)\). Consequently, \(H'=2(\Ree a)|b|^2-[2(\Ree a)|b|^2+2c\Ree(b^2)]+2c\Ree(b^2)=0\). The terms containing the twist cancel in the derivative of \(|a|^2\), while the exchange between the electromagnetic and scalar contributions cancels between the last two terms of \(H\). Thus the initial values of \(a,b,c\) fix one constant throughout the integration. This is the affine potential-space energy of the geodesic, see Ref. \cite{Bixano:2026duf}. 

Conserved energy sets the norm of the tangent vector in potential space. At the normalized asymptotic point, Eq \eqref{eq:charges} relates this norm to the physical charges, whereas \(k=HK[\s]\) fixes how it enters the spatial metric. Any nontrivial geodesic with \(H=0\) is null. The branches with positive and negative energy are likewise physically meaningful: their regularity is ensured by the consistency between the conserved quantities and the selected harmonic function, together with the boundary behavior of the reconstructed fields.

Let the dual harmonic function \(\schdual\) and the \(k\)-Weyl functional \(K[\s]\) defined by
\begin{equation}
     \begin{aligned}
     D\schdual&=-\rho(\partial_z\s,-\partial_\rho\s),\\
     K_{,\rho}&=\rho(\s_{,\rho}^2-\s_{,z}^2),&
     K_{,z}&=2\rho\s_{,\rho}\s_{,z}.
     \end{aligned}
     \label{eq:dualK}
\end{equation}
Both expressions are integrable since \(\s\) is harmonic. For instance, \(\partial_zK_{,\rho}-\partial_\rho K_{,z}=-2\rho\s_{,z}\rho^{-1}D\!\cdot(\rho D\s)=0\). With the normalized initial data written as
\begin{equation}
     \begin{gathered}
     a(0)=\frac{l_1+i l_2}{L},\qquad
     b(0)=\frac{-q_0+i p_0}{L},\\
     c(0)=-\frac{\al d_0}{L},\\
     HL^2=l_1^2+l_2^2+d_0^2-q_0^2-p_0^2,
     \end{gathered}
     \label{eq:charges}
\end{equation}
the remaining quadratures take the form
\begin{equation}
     k=HK[\s],\qquad \omega=\frac{2l_2}{L}\schdual,\qquad
     A_\varphi=-\omega A_t-\frac{p_0}{L}\schdual.
     \label{eq:quadratures}
\end{equation}
We have fixed the additive constants by choosing a symmetric local angular gauge. The angular reconstruction can be read off as follows: combining Eq \eqref{eq:abc} with Eq \eqref{eq:reconstruct} yields the conserved quantities \(\Imm a/f=l_2/L\) and \(\Imm b/(\kappa\sqrt f)-(l_2/L)\psi=p_0/L\). The twist equation then integrates directly to the stated \(\omega\). Moreover, differentiating the given expression for \(A_\varphi\) produces \(DA_\varphi=-\omega DA_t-[2l_2A_t+p_0]D\schdual/L\), which is precisely the magnetic reconstruction equation. A full account of the procedure is provided in \cite{Bixano:2026duf}.

\subsection{Spheroidal normalization and physical charges}\label{sec:coordinates}
To provide a physical interpretation, we work in prolate spheroidal coordinates defined by \(\rho=L\sqrt{(x^2-1)(1-y^2)}\) and \(z=Lxy\), where \(L>0\), \(x>1\), and \(-1<y<1\). In these variables, the axisymmetric Laplace equation takes the form \(\partial_x[(x^2-1)\s_{,x}]+\partial_y[(1-y^2)\s_{,y}]=0\), while the flat \((\rho,z)\) line element becomes
\begin{equation}
     \begin{split}
     \dd\rho^2+\dd z^2={}&L^2(x^2-y^2)
     \left[\frac{\dd x^2}{x^2-1}+\frac{\dd y^2}{1-y^2}\right].
     \end{split}
     \label{eq:spheroidalbase}
\end{equation}
Using a monopole harmonic function, i.e. \(\s=-x^{-1}+O(x^{-2})\), the quantities appearing in Eq \eqref{eq:charges} can be identified with the physical mass, NUT, electric, magnetic, and scalar charges. If one instead uses a general harmonic lacking this normalization, these quantities should be regarded simply as parameters.

Using the monopole normalization and considering a monopole harmonic function presented in \eqref{eq:Schharmonic}, \eqref{eq:reconstruct} yields
\begin{equation}
     \begin{aligned}
     f&=1-\frac{2l_1}{Lx}+O(x^{-2}),&
     A_t&=-\frac{q_0}{Lx}+O(x^{-2}),\\
     \phi&=\frac{d_0}{Lx}+O(x^{-2}),&
     \chi&=-\frac{2p_0}{Lx}+O(x^{-2}).
     \end{aligned}
     \label{eq:asymptotics}
\end{equation}
Accordingly, \(l_1\) is identified with the geometric mass, \(q_0\) with the geometric electric charge,  \(p_0\) with the geometric magnetic charge, and \(d_0\) with the geometric scalar charge. The normalizations of \(q_0\) and \(p_0\) follow from the electric flux \((4\pi)^{-1}\int e^{-2\al\phi}\,{}^\star F\) and the magnetic flux \((4\pi)^{-1}\int F\). The constant \(l_2\) is determined by the asymptotic angular twist and is understood as the geometric NUT parameter. In all subsequent asymptotic evaluations we work with \(x\), keeping \(y\) fixed.

Only in presenting the final radial form do we switch to Boyer--Lindquist-type coordinates, using
\begin{equation}
     Lx=r-l_1,\qquad y=\cos\theta.
     \label{eq:BL}
\end{equation}
This choice enforces a standard mass-centered convention for the radial coordinate. Its asymptotic interpretation is read off from \(\rho^2/f=[(Lx)^2+2l_1Lx+O(1)](1-y^2)\): the selected shift ensures that the dominant angular term is \(r^2(1-y^2)\), while any finite NUT contribution only appears at subleading order. For later convenience, define the geometric combinations \(\Delta=(r-l_1)^2-L^2\) and \(\Sigma=(r-l_1)^2-L^2\cos^2\theta\), so that \(\rho^2=\Delta\sin^2\theta\) and \(\dd\rho^2+\dd z^2=\Sigma(\dd r^2/\Delta+\dd\theta^2)\).
We will use these expressions only when they render the reconstructed metric more compact. Finally, Eq \eqref{eq:charges} clarifies what is captured in \(L\): for \(H=1\) it reduces to the full nonextremality combination of the charges, for \(H\ne 1\) the overall factor \(H\) must be retained, and for \(H=0\) the same relation turns into a charge-balance condition rather than an equation fixing \(L\), as in the extremal black-hole configurations discussed below.

\section{Integration and identification of one-harmonic branches}\label{sec:one}

\subsection{Static vacuum: the arbitrary harmonic and the Schwarzschild case}

\subsubsection{Schwarzschild branch in terms of an arbitrary harmonic function}
To arrive at a static setup, we take \(a\) to be a real function, so its imaginary part vanishes and thus \(\omega=0\). Eliminating the Maxwell sector amounts to imposing \(b=0\), and turning off the scalar field corresponds to \(\phi=0\). We further enforce the pure-vacuum limit by also choosing \(\al=0\) and \(c=0\). With these choices, the only remaining field equation is \(a'=0\), which implies \(a=h\) for an arbitrary constant \(h\in\mathbb R\). From the first reconstruction relation one then finds \(\ln f=2h\s\), while \(A_t'=\chi'=0\). The energy is \(H=h^2\), and consequently \(k=h^2K[\s]\). Keeping the harmonic function arbitrary, the full solution is
\begin{equation}
     \begin{aligned}
     \dd s^2={}&-e^{2h\s}\dd t^2+
     e^{-2h\s}\left[e^{2h^2K[\s]}(\dd\rho^2+\dd z^2)
     +\rho^2\dd\varphi^2\right],\\
     \mathcal A={}&0,\qquad \phi=0.
     \end{aligned}
     \label{eq:vacuumharmonic}
\end{equation}
The parameters satisfy \(l_1=hL\) and \(l_2=q_0=p_0=d_0=0\). Consequently, the Hamiltonian condition becomes \(H=l_1^2/L^2=h^2\). If, on the other hand, \(\s\) includes higher multipole terms, then Eq \eqref{eq:vacuumharmonic} represents the associated static Weyl vacuum solution.

\subsubsection{Spheroidal coordinates and the physical role of \texorpdfstring{\(h\)}{h}}

We next adopt the Schwarzschild harmonic function together with the corresponding \(k\)-quadrature \(K[\s]\) and the associated dual harmonic function \(\schdual\),
\begin{equation}
     \begin{gathered}
     \s=\frac12\ln\frac{x-1}{x+1},\qquad \schdual=Ly,\\
     K[\s]=\frac12\ln\frac{x^2-1}{x^2-y^2}.
     \end{gathered}
     \label{eq:Schharmonic}
\end{equation}
The harmonic equation is obtained directly from \((x^2-1)\s_{,x}=1\), since \(\s(x)\) depends only on the coordinate \(x\), and the dual function follows from Eq \eqref{eq:dualK}. The asymptotic expansion reads \(\s=-x^{-1}-(3x^3)^{-1}+O(x^{-5})\).

In Eq \eqref{eq:vacuumharmonic}, define \(f_h=[(x-1)/(x+1)]^h\) and apply \eqref{eq:spheroidalbase}. This yields the Zipoy--Voorhees class \cite{Voorhees:1970ywo,Quevedo:2012ttw}, in which the redshift is multiplied by \(h\) while the meridional Weyl factor is multiplied by \(h^2\). Both rescalings have physical significance. Along the upper axis, the vacuum Ernst combination admits the series
\begin{equation}
     \begin{aligned}
     \frac{1-f_h}{1+f_h}
     &=\tanh\!\left[h\operatorname{arctanh}(x^{-1})\right]\\
     &=\frac{h}{x}+\frac{h(1-h^2)}{3x^3}+O(x^{-5}).
     \end{aligned}
     \label{eq:ZVmoments}
\end{equation}
The centered mass moments are \(\mathcal M_0=hL=l_1\) and \(\mathcal M_2=h(1-h^2)L^3/3\), consistent with the invariant moments of the \(q\)-metric \cite{Quevedo:2012ttw}. Hence, varying \(h\) while keeping the physical mass fixed necessarily alters the quadrupole moment.

For reference, Eq. \eqref{eq:BL} yields the full deformed configuration,
\begin{equation}
     \begin{aligned}
     f_h&=\left(\frac{r-l_1-L}{r-l_1+L}\right)^h,
     \qquad l_1=hL,\\
     \dd s^2={}&-f_h\dd t^2+f_h^{-1}
     \left(\frac{\Delta}{\Sigma}\right)^{h^2}
     \Sigma\left(\frac{\dd r^2}{\Delta}+\dd\theta^2\right)\\
     &+f_h^{-1}\Delta\sin^2\theta\dd\varphi^2,\\
     \mathcal A={}&0,\qquad\phi=0.
     \end{aligned}
     \label{eq:ZVradial}
\end{equation}
In general, the surface \(x=1\) does not correspond to a regular black-hole horizon. When \(h>0\), the density of the angular area (apart from the endpoints) scales as \(\sqrt{g_{yy}g_{\varphi\varphi}} \propto (x-1)^{(h-1)^2/2}\). A finite nonvanishing angular metric on this surface occurs only for \(h=1\), the curvature singularities present for the other positive-mass cases are discussed in \cite{Quevedo:2012ttw}.

\paragraph{Schwarzschild solution.}
Setting \(h=1\) implies \(H=1\) and \(L=l_1>0\). The imposed coordinate condition becomes \(Lx=r-l_1\), so \(f_1=(x-1)/(x+1)=1-2l_1/r\) and \(\Delta=r(r-2l_1)\). In Eq. \eqref{eq:ZVradial} the prefactor \(\Sigma\) drops out against the Weyl quadrature, which yields
\begin{equation}
     \begin{aligned}
     \dd s^2={}&-\left(1-\frac{2l_1}{r}\right)\dd t^2
     +\frac{\dd r^2}{1-2l_1/r}+r^2\dd\Omega_2^2,\\
     \mathcal A={}&0,\qquad\phi=0,
     \quad \dd\Omega_2^2=\dd\theta^2+\sin^2\theta\dd\varphi^2.
     \end{aligned}
     \label{eq:Schwarzschild}
\end{equation}
In this parametrization, \(r\) coincides with the areal radius, and \(r=2l_1\) corresponds to a smooth, nonextremal horizon. These features arise from the cancellation specific to \(h=1\). Taking \(h=0\) reproduces flat Minkowski spacetime. For \(L>0\), choosing \(h=-1\) instead leads to the negative-mass Schwarzschild exterior, which has no black-hole horizon at any positive radius and therefore should be excluded from the positive-mass interpretation discussed above.

In the description of the potential-space, \(h\) fixes the gravitational geodesic energy \(H=h^2\). With the Schwarzschild harmonic held fixed, the positive-mass horizon condition selects \(h=1\), and therefore \(H=1\). The same choice removes the independent quadrupolar deformation and fixes \(L=l_1\).

\subsection{Static scalar-vacuum solutions and the Fisher--Janis--Newman--Winicour case}

\subsubsection{Branch in terms of an arbitrary harmonic function}

To keep a nontrivial scalar sector while maintaining a static configuration and eliminating the electromagnetic field, we again choose \(a(\s)\) to be real and impose \(b(\s)=0\). The first and third relations in \eqref{eq:abc} then imply \(a'=c'=0\). These integration constants are determined directly from the initial conditions, yielding \(a=l_1/L\) and \(c=-\al d_0/L\). Hence the scalar charge stays decoupled from the mass. Upon integrating \eqref{eq:reconstruct}, one finds \(f=\exp(2l_1\s/L)\), \(\phi=-d_0\s/L\), and \(\kappa=\exp(\al d_0\s/L)\). With the asymptotic normalization already fixed, the electromagnetic potentials are identically zero. The most general harmonic solution in this class is therefore
\begin{equation}
    \begin{aligned}
     \dd s^2={}&-e^{2l_1\s/L}\dd t^2\\
     &+e^{-2l_1\s/L}\left[e^{2HK[\s]}(\dd\rho^2+\dd z^2)
     +\rho^2\dd\varphi^2\right],\\
     \mathcal A={}&0,\qquad \phi=-\frac{d_0}{L}\s,
     \qquad H=\frac{l_1^2+d_0^2}{L^2}.
    \end{aligned}
    \label{eq:scalarharmonic}
\end{equation}
The scalar modifies both its own configuration and the Weyl function via \(H\), reflecting the positive contribution of the scalar to the energy. When \(\al=0\), Eq \eqref{eq:scalarharmonic} is recovered by combining the independent free scalar with the energy obtained earlier. In that limit the metric and scalar retain the same functional forms, since the electromagnetic sector is absent. This also clarifies why the scalar charge remains physically relevant even for vanishing dilaton coupling.

\subsubsection{Spheroidal configuration and the scalar charge}

We now choose \eqref{eq:Schharmonic}. The spherical scalar-vacuum case has \(H=1\), which fixes \(L^2=l_1^2+d_0^2\). Substitution of \(k=K[\s]\) and Eq \eqref{eq:spheroidalbase} then gives
\begin{equation}
    \begin{aligned}
     \dd s^2={}&-f\dd t^2+\frac{L^2}{f}\dd x^2\\
     &+\frac{L^2(x^2-1)}{f}
     \left[\frac{\dd y^2}{1-y^2}+(1-y^2)\dd\varphi^2\right],\\
     f={}&\left(\frac{x-1}{x+1}\right)^{l_1/L},\qquad
     \mathcal A=0,\\
     \phi={}&-\frac{d_0}{2L}\ln\frac{x-1}{x+1}.
    \end{aligned}
    \label{eq:FisherSph}
\end{equation}
This corresponds to the Fisher--Janis--Newman--Winicour solution \cite{Fisher:1948yn,Janis:1968zz}. Its expansion at large \(x\) reproduces the mass and scalar charge given in Eq \eqref{eq:asymptotics}. Using \(Lx=r-l_1\), one obtains the following final expression in the radial coordinate:
\begin{equation}
    \begin{aligned}
     f={}&\left(\frac{r-l_1-L}{r-l_1+L}\right)^{l_1/L},\\
     \dd s^2={}&-f\dd t^2+f^{-1}\dd r^2+
     \frac{\Delta}{f}\dd\Omega_2^2,\qquad \mathcal A=0,\\
     \phi={}&-\frac{d_0}{2L}\ln\frac{r-l_1-L}{r-l_1+L},
     \qquad L^2=l_1^2+d_0^2.
    \end{aligned}
    \label{eq:FisherRadial}
\end{equation}
For \(L>0\), \(l_1>0\), and a nonvanishing scalar charge, the parameters satisfy \(0<l_1/L<1\). Near \(x=1\), the angular prefactor in \eqref{eq:FisherSph} scales as \((x-1)^{1-l_1/L}\), implying that the area shrinks to zero. A more immediate curvature diagnostic follows from the trace of Einstein’s equation,
\(R=2d_0^2f/[L^4(x^2-1)^2]\), which diverges at \(x=1\) in this interval of \(l_1/L\). Hence the scalar charge converts the Schwarzschild horizon into the familiar scalar (naked) singularity. In the special case \(d_0=0\), the positive-mass branch yields \(L=l_1\), and the metric, four-potential, and scalar field simultaneously reduce to Eq \eqref{eq:Schwarzschild}. The more general option \(H\ne1\) in \eqref{eq:scalarharmonic} is still possible, for the same Schwarzschild harmonic it preserves an extra axisymmetric deformation.

Setting \(H=1\) determines the total gravitational-plus-scalar contribution, \(l_1^2+d_0^2=L^2\), but leaves their individual shares unconstrained. The presence of the scalar singularity illustrates why the energy must be treated jointly with the scalar configuration: with the same normalization one can obtain either a regular Schwarzschild horizon or a singular Fisher boundary.

\subsection{Static electrovacuum and dyonic Reissner--Nordstr\"om}

\subsubsection{Branch in terms of an arbitrary harmonic function with independent electric and magnetic charges}

To introduce an electromagnetic field without changing the underlying geometry, we keep \(a(\s)\) real and permit \(b(\s)\) to be complex. We set \(\al=0\) and \(\phi(\s)=0\), which implies \(c(\s)=0\). The field equations then reduce to \(a'=|b|^2\) and \(b'=ab\). Wherever the real part of \(b(\s)\) is nonzero, the ratio between the imaginary and real components of \(b(\s)\) remains constant. Consequently, the electric and magnetic charges share the same radial dependence. Their independent initial values are already determined by Eq \eqref{eq:charges}.

The energy relation becomes \(a'=a^2-H\). To integrate, we use \(a=-\mathcal D'/\mathcal D\) with \(\mathcal D(0)=1\). Inserting this ansatz turns the equation into \(\mathcal D''=H\mathcal D\), and the initial condition is \(\mathcal D'(0)=-l_1/L\). When \(H=h^2>0\), a normalized solution is
\begin{equation}
    \begin{aligned}
     \mathcal D(\s)&=\cosh(h\s)-\frac{l_1}{hL}\sinh(h\s),\\
     a&=-\frac{\mathcal D'}{\mathcal D},\qquad
     b=\frac{-q_0+i p_0}{L\mathcal D},\qquad c=0,\\
     h^2L^2&=l_1^2-q_0^2-p_0^2.
    \end{aligned}
    \label{eq:staticEMabc}
\end{equation}
Here \(\mathcal D(\s)\) serves as the shared denominator for the integrated fields. It is determined by the charges, not introduced as an additional independent potential. On the real branch attached to spatial infinity we choose \(\mathcal D>0\). The first integral \(\mathcal D'^2-h^2\mathcal D^2=(q_0^2+p_0^2)/L^2\) confirms \(a'=|b|^2\), and this remains true in the purely electric or purely magnetic limits.

The redshift factor integrates to \(f=\mathcal D^{-2}\). In addition, \(A_t'=q_0/(L\mathcal D^2)\) and \(\chi'=2p_0/(L\mathcal D^2)\). Their antiderivatives are obtained using \([\sinh(h\s)/(h\mathcal D)]'=\mathcal D^{-2}\). Hence, the three physical fields, prior to selecting a specific harmonic, read
\begin{equation}
    \begin{aligned}
     \dd s^2={}&-\mathcal D^{-2}\dd t^2+
     \mathcal D^2\left[e^{2h^2K[\s]}(\dd\rho^2+\dd z^2)
     +\rho^2\dd\varphi^2\right],\\
     \mathcal A={}&\frac{q_0\sinh(h\s)}{hL\mathcal D}\dd t
     -\frac{p_0}{L}\schdual\dd\varphi,\qquad \phi=0.
    \end{aligned}
    \label{eq:staticEMharmonic}
\end{equation}
The magnetic contribution to the four-potential is recovered via the dual quadrature. In particular, obtaining \(\chi\) by integration does not necessitate altering the angular gauge that was already chosen in Eq \eqref{eq:quadratures}. The metric only involves the combination \(q_0^2+p_0^2\), while the two components of the four-potential separately encode the two charges.

\subsubsection{The charged spheroidal family and the regular case}

Using \eqref{eq:Schharmonic} and imposing the same coordinate choice \(Lx=r-l_1\) as in the vacuum case, and evaluating \(\mathcal D\) at
\(\s=\tfrac12\ln[(r-l_1-L)/(r-l_1+L)]\), one obtains the complete deformed configuration
\begin{equation}
    \begin{aligned}
     \dd s^2={}&-\mathcal D^{-2}\dd t^2+
     \mathcal D^2\left(\frac{\Delta}{\Sigma}\right)^{h^2}
     \Sigma\left(\frac{\dd r^2}{\Delta}+\dd\theta^2\right)\\
     &+\mathcal D^2\Delta\sin^2\theta\dd\varphi^2,\\
     \mathcal A={}&\frac{q_0\sinh(h\s)}{hL\mathcal D}\dd t
     -p_0\cos\theta\dd\varphi,\qquad \phi=0.
    \end{aligned}
    \label{eq:chargedZVradial}
\end{equation}
This represents the dyonic extension of the charged Zipoy--Voorhees class \cite{Gurtug:2021noy}. For the purely electric representative, matching to the hyperbolic parametrization of that solution shows that the distortion parameter is \(h\), while a fixed Maxwell duality rotation independently sets \(q_0\) and \(p_0\). For \(h\neq 1\), the radial redshift profile and the meridional power both get modified, in direct analogy with the vacuum case.

When \(l_1>0\) and \(h>0\), the denominator behaves near \(x=1\) as \(\mathcal D\sim[1+l_1/(hL)]e^{-h\s}/2\). Hence \(f\) shares the same dominant scaling \((x-1)^h\) as the vacuum redshift factor. Using in addition \(k=h^2K\), the angular area density scales as \((x-1)^{(h-1)^2/2}\). Requiring the horizon area to be finite and nonzero fixes \(h=1\) on this positive-mass branch, providing a physical constraint on the otherwise arbitrary integration constants.

\paragraph{Dyonic Reissner--Nordstr\"om solution.}
Setting \(h=1\), the energy becomes \(H=1\), and \(L^2=l_1^2-q_0^2-p_0^2>0\). Using \(\cosh\s=x/\sqrt{x^2-1}\) and \(\sinh\s=-1/\sqrt{x^2-1}\), one obtains \(\mathcal D=(Lx+l_1)/(L\sqrt{x^2-1})\). It follows that \(f=L^2(x^2-1)/(Lx+l_1)^2\), \(A_t=-q_0/(Lx+l_1)\), and \(A_\varphi=-p_0y\). Since the Weyl factor drops out of the meridional sector of the metric, Eq \eqref{eq:BL} yields
\begin{equation}
    \begin{aligned}
     f={}&1-\frac{2l_1}{r}+\frac{q_0^2+p_0^2}{r^2},\\
     \dd s^2={}&-f\dd t^2+f^{-1}\dd r^2+r^2\dd\Omega_2^2,\\
     \mathcal A={}&-\frac{q_0}{r}\dd t-p_0\cos\theta\dd\varphi,
     \qquad \phi=0.
    \end{aligned}
    \label{eq:RN}
\end{equation}
The function \(f\) has two zeros at \(r=l_1\pm L\). The outer horizon therefore has area \(4\pi(l_1+L)^2\), and evaluating \(\tfrac12 f'(r)\) at this radius gives \(L/(l_1+L)^2\), i.e. the surface gravity. These relations make explicit that the electric and magnetic charges decrease the nonextremality parameter while leaving the mass \(l_1\) unchanged.

\subsubsection{Zero and negative energy}\label{sec:zero-negative-energy}

The linear expression for \(\mathcal D\) preserves the sign structure of \(H\). When \(H=0\), \(\mathcal D=1-l_1\s/L\), and the factor \(\sinh(h\s)/h\) appearing in the four-potential reduces to \(\s\). For \(H<0\), the denominator remains real and is given by \(\mathcal D=\cos(\sqrt{-H}\s)- l_1\sin(\sqrt{-H}\s)/(L\sqrt{-H})\). These relations correspond to integrations of the same arbitrary-harmonic field equations. To single out the spherical configurations, one chooses the harmonic and its spatial quadrature together with the energy.

In the extremal spherically symmetric limit, set \(H=0\), \(l_1^2=q_0^2+p_0^2\), and choose the Coulomb harmonic \(\s=-1/x\), with \(\rho=Lx\sqrt{1-y^2}\), \(z=Lxy\), and \(\schdual=Ly\). For this metric one finds \(k=0\) and \(f=[1+l_1/(Lx)]^{-2}\). Using the redefinition \(Lx=r-l_1\) reproduces Eq \eqref{eq:RN} with \(f=(1-l_1/r)^2\), along with the same four-potential and \(\phi=0\). The horizon is then degenerate and located at \(r=l_1\). In this static zero-energy class, taking more general harmonic functions yields the Majumdar--Papapetrou geometries after a constant electromagnetic duality rotation \cite{Majumdar:1947eu,Papaetrou:1947ib}.

When \(q_0^2+p_0^2>l_1^2\), take \(H=-1\) and define \(L^2=q_0^2+p_0^2-l_1^2\). In the oblate form one may use \(\rho=L\sqrt{(x^2+1)(1-y^2)}\), \(z=Lxy\), \(\s=\arctan x-\pi/2\), and \(\schdual=Ly\), with \(K=\tfrac12\ln[(x^2+y^2)/(x^2+1)]\). In this case \(k=-K\), and, after the identification \(Lx=r-l_1\), the spherical fields once more collapse to Eq \eqref{eq:RN}. The associated redshift function has no real zero. This construction therefore keeps the superextremal RN case explicit while maintaining \(L\in\mathbb{R}\). The coordinate ranges are selected to avoid \(r=0\), where both the curvature and the electromagnetic field diverge.

The relation \(\mathcal D''=H\mathcal D\) encodes the three energy regimes via hyperbolic, linear, and trigonometric integrations. Together with their associated spherical-harmonic implementations, these reproduce the nonextremal, extremal, and superextremal RN solutions. Hence, charge balance and spatial structure are linked by the same conserved energy in potential space.

\subsection{Complex electrovacuum and the charged Taub--NUT solution}

\subsubsection{Exact integration with a complex gravitational coefficient}

To keep a nonzero twist, we now permit \(a(\s)\) to take complex values, while still imposing \(\al=0\), \(c(\s)=0\), and \(\phi=0\). The electromagnetic coefficient \(b(\s)\) is likewise complex, so the two charges remain independent. As before, the energy constraint reduces the first equation to \(a'=a^2-H\). The associated normalized complex denominator can be written as
\begin{equation}
    \begin{aligned}
     \mathcal D={}&\cosh(h\s)
     -\frac{l_1+i l_2}{hL}\sinh(h\s),\\
     a={}&-\frac{\mathcal D'}{\mathcal D},\qquad
     b=\frac{-q_0+i p_0}{L}
     \frac{\mathcal D}{\overline{\mathcal D}\,|\mathcal D|},\\
     c={}&0,\qquad
     H=h^2=\frac{l_1^2+l_2^2-q_0^2-p_0^2}{L^2}.
    \end{aligned}
    \label{eq:complexEMabc}
\end{equation}
The cases \(H=0\) and \(H<0\) are again described by the same linear or trigonometric forms as above. To verify the solution directly when \(H>0\), set \(\mathcal D=|\mathcal D|e^{i\arg\mathcal D}\). Then \(\Ree a=-(\ln|\mathcal D|)'\) and \(\Imm a=-(\arg\mathcal D)'\), and differentiating the formula for \(b\) yields \(b'/b=\Ree a-2i\Imm a\), which matches its field equation. The relation \(\mathcal D'\overline{\mathcal D}'- h^2|\mathcal D|^2=(q_0^2+p_0^2)/L^2\) confirms the energy condition. Moreover, the imaginary Wronskian implies \(\Imm a=(l_2/L)|\mathcal D|^{-2}\), reproducing the twist integral previously obtained from the Noether charge.

Reconstructing the electromagnetic potentials is especially straightforward when one considers the specific combination \((A_t+i\chi/2)'=-\sqrt f\,\bar b\). With \(f=|\mathcal D|^{-2}\), the right-hand side can be rewritten as \((q_0+i p_0)/(L\mathcal D^2)\). Invoking the same Wronskian identity as in the static branch then yields
\begin{equation}
    \begin{aligned}
     A_t&=\Ree\!\left[
     \frac{(q_0+i p_0)\sinh(h\s)}{hL\mathcal D}\right],\\
     \chi&=2\Imm\!\left[
     \frac{(q_0+i p_0)\sinh(h\s)}{hL\mathcal D}\right].
    \end{aligned}
    \label{eq:complexEMpotentials}
\end{equation}
At \(\s=0\), both potentials are zero. Their real and imaginary components encode the two independent electromagnetic charges, whereas their radial mixing is determined by \(l_2\).

The general solution for an arbitrary harmonic is
\begin{equation}
    \begin{aligned}
     \dd s^2={}&-\frac{1}{|\mathcal D|^2}
     \left(\dd t-\frac{2l_2}{L}\schdual\dd\varphi\right)^2\\
     &+|\mathcal D|^2\left[e^{2HK[\s]}(\dd\rho^2+\dd z^2)
     +\rho^2\dd\varphi^2\right],\\
     \mathcal A={}&A_t\left(\dd t-\frac{2l_2}{L}\schdual\dd\varphi\right)
     -\frac{p_0}{L}\schdual\dd\varphi,
     \qquad \phi=0,
    \end{aligned}
    \label{eq:complexEMharmonic}
\end{equation}
where \(A_t\) is written explicitly in Eq \eqref{eq:complexEMpotentials}. As \(x\to\infty\), one has \(\mathcal D=1-(l_1+i l_2)\s/L+O(\s^2)\), and therefore the conventions for the mass, NUT parameter, and electromagnetic charges agree with Eq \eqref{eq:charges}. 

\subsubsection{Spheroidal coordinates and the Brill case}

For the Schwarzschild harmonic, setting \(k=HK\) leaves the meridional deformation unchanged. Taking \(H=1\) (equivalently \(h=1\)) yields \(\mathcal D=(Lx+l_1+i l_2)/(L\sqrt{x^2-1})\). It then follows that \(f=L^2(x^2-1)/[(Lx+l_1)^2+l_2^2]\), \(\omega=2l_2y\), and \(A_t=-[q_0(Lx+l_1)+p_0l_2]/[(Lx+l_1)^2+l_2^2]\).
The radial and angular metric coefficients are obtained by inserting \(k=K\) into Eq \eqref{eq:spheroidalbase}. Using \(Lx=r-l_1\) and \(y=\cos\theta\), one arrives at the following three fields:
\begin{equation}
    \begin{aligned}
     f={}&\frac{\Delta}{r^2+l_2^2},\qquad
     L^2=l_1^2+l_2^2-q_0^2-p_0^2,\\
     \dd s^2={}&-f(\dd t-2l_2\cos\theta\dd\varphi)^2
     +f^{-1}\dd r^2+(r^2+l_2^2)\dd\Omega_2^2,\\
     \mathcal A={}&-\frac{q_0r+p_0l_2}{r^2+l_2^2}
     (\dd t-2l_2\cos\theta\dd\varphi)
     -p_0\cos\theta\dd\varphi,\\
     \phi={}&0.
    \end{aligned}
    \label{eq:Brill}
\end{equation}
This corresponds to the dyonically charged Taub--NUT solution originally discussed by Brill \cite{PhysRev.133.B845,Clement:2015aka}. The matching applies to both the metric and the electromagnetic sector. In particular, one has \(F_{rt}=[q_0(r^2-l_2^2)+2p_0l_2r]/(r^2+l_2^2)^2\), with leading asymptotic behavior \(q_0/r^2\). The asymptotic magnetic term is set by the potential \(-p_0\cos\theta\dd\varphi\). Hence, taking complex \(a,b\) with \(c=0\) reproduces a familiar electrovacuum family, within these conventions, their phases encode the NUT contribution and the mixing between gravitational and electromagnetic parameters.

The complex gravitational coefficient supplements the geodesic information with the conserved twist charge. It contributes to the energy via \(l_2^2\), whereas the electromagnetic part contributes as \(-q_0^2-p_0^2\). The condition that determines \(L\) thus comes straight from the geodesic energy, and the individual Noether integrals preserve the physical separation between the NUT parameter and the electromagnetic charges.

The outer root occurs at \(r=l_1+L\). The corresponding local horizon area is \(4\pi[(l_1+L)^2+l_2^2]\), and the surface gravity is \(L/[(l_1+L)^2+l_2^2]\). The dominant angular twist is \(2l_2y\), representing the NUT contribution. By contrast, a Kerr angular-momentum term has a different angular profile and falls off at infinity. The additive constant in \(\omega\) can be chosen, within a local angular gauge, to locate the Misner string on either axis, its global handling is part of the standard Taub--NUT geometry \cite{Clement:2015aka}.

Setting \(q_0=p_0=0\), the same computation yields the vacuum Taub--NUT solution, with \(L^2=l_1^2+l_2^2\), the metric as in Eq \eqref{eq:Brill}, \(\mathcal A=0\), and \(\phi=0\). When \(l_2=0\), Eq \eqref{eq:Brill} specializes to the full RN configuration. Keeping \(H=h^2\ne1\) together with the Schwarzschild harmonic, the functions in Eq \eqref{eq:complexEMharmonic} reproduce the stationary charged Zipoy--Voorhees class. A comparison to the established stationary solution \cite{Halilsoy:2022pzr}, along with a parameter map on a shared real domain, is presented in Appendix \ref{app:chargedZV}. This, in turn, fixes the physical meaning of \(l_1,l_2,q_0,p_0\) consistently throughout the integration.

\paragraph{The constant-scalar restriction at nonzero coupling.}
The significance of setting \(\al=0\) on this branch follows immediately from the scalar equation. When \(\al\neq 0\) with \(c=0\), the equation enforces \((\Ree b)^2=(\Imm b)^2\). Taking a derivative of this relation and using the evolution equation for \(b\) yields \(0=8(\Imm a)(\Ree b)(\Imm b)\). Along a nontrivial electromagnetic branch, both electromagnetic factors in this product are nonzero, which forces \(\Imm a=0\). The remaining constant-scalar configuration is time-independent and satisfies \(q_0^2=p_0^2\). Its invariant \(F_{\mu\nu}F^{\mu\nu}\) is zero, in agreement with the scalar field equation. This reproduces the balanced RN solution for arbitrary continuous \(\al\), whereas the generic charged NUT solution is described by the electrovacuum system integrated earlier.

\subsection{The real electric dilatonic branch and Gibbons--Maeda solution}

\subsubsection{Integration with an independent scalar charge}

To ensure a fixed geometry and an electromagnetic field that is purely electric, we choose \(a(\s),b(\s),c(\s)\) to be real, with \(b\ne0\). We begin with the case \(\al\ne0\). The system consists of \(a'=b^2\), \(b'=(a+c)b\), and \(c'=\al^2b^2\). Comparing the first and third equations yields \((c-\al^2a)'=0\). Its constant is determined by the physical initial conditions, which imply \(c-\al^2a=-\al(d_0+\al l_1)/L\). This constraint preserves an independent scalar charge and reduces the remaining integration problem to a single electromagnetic denominator.

Next, differentiating \(a+c\) gives \((a+c)'=(1+\al^2)b^2\), together with \(b'=(a+c)b\). Hence \([(a+c)^2-(1+\al^2)b^2]'=0\). Let the associated constant be \(\lambda^2\), and set \(a+c=-(\ln\mathcal D_e)'\). From the equation for \(b\) one then obtains \(b=-q_0/(L\mathcal D_e)\). The last step is the linear ODE \(\mathcal D_e''=\lambda^2\mathcal D_e\), with normalized conditions we use \(a(0)+c(0)=(l_1-\al d_0)/L\). Consequently,
\begin{equation}
    \begin{aligned}
     \lambda^2&=\frac{(l_1-\al d_0)^2-(1+\al^2)q_0^2}{L^2},\\
     \mathcal D_e&=\cosh(\lambda\s)
     -\frac{l_1-\al d_0}{\lambda L}\sinh(\lambda\s),\\
     a&=\frac{-\mathcal D_e'/\mathcal D_e+
     \al(d_0+\al l_1)/L}{1+\al^2},\\
     b&=-\frac{q_0}{L\mathcal D_e},\qquad
     c=\al^2a-\frac{\al(d_0+\al l_1)}{L}.
    \end{aligned}
    \label{eq:GMabc}
\end{equation}
The parameter \(\lambda^2\) is therefore fixed by the mass, the electric charge, and the scalar charge. When it vanishes, one finds \(\mathcal D_e=1-(l_1-\al d_0)\s/L\). If \(\lambda^2\) is negative, the same real solution follows by substituting the hyperbolic functions with their trigonometric counterparts, as done in the electrovacuum integration. In each case, the complete local real solution for this branch is preserved.

To recover the metric, we integrate \(2a\) in Eq \eqref{eq:GMabc}. Performing the integration of \(c/\al\) simultaneously yields the scalar field. The electric equation can be reduced further since \(f/\kappa^2=\mathcal D_e^{-2}\), which implies \(A_t'=q_0/(L\mathcal D_e^2)\). An antiderivative is determined using the identity \([\sinh(\lambda\s)/(\lambda\mathcal D_e)]' =\mathcal D_e^{-2}\). The complete solution for an arbitrary harmonic is
\begin{equation}
    \begin{aligned}
     f={}&\exp\!\left[
     \frac{2\al(d_0+\al l_1)}{(1+\al^2)L}\s\right]
     \mathcal D_e^{-2/(1+\al^2)},\\
     \dd s^2={}&-f\dd t^2+
     f^{-1}\left[e^{2HK[\s]}(\dd\rho^2+\dd z^2)
     +\rho^2\dd\varphi^2\right],\\
     \mathcal A={}&\frac{q_0\sinh(\lambda\s)}
     {\lambda L\mathcal D_e}\dd t,\\
     \phi={}&-\frac{\al}{1+\al^2}\ln\mathcal D_e
     -\frac{d_0+\al l_1}{(1+\al^2)L}\s,\\
     H={}&\frac{l_1^2+d_0^2-q_0^2}{L^2}
     =\frac{\lambda^2+(d_0+\al l_1)^2/L^2}{1+\al^2}.
    \end{aligned}
    \label{eq:GMharmonic}
\end{equation}
In this case \(l_2=p_0=0\), and the physical branch attached to the selected asymptotic region satisfies \(\mathcal D_e>0\). At \(\s=0\), the scalar derivative in \eqref{eq:GMharmonic} equals \(-d_0/L\), the two terms add up to exactly the required scalar charge. An analogous check yields \(f'(0)=2l_1/L\) and \(A_t'(0)=q_0/L\). This establishes the correct physical normalization prior to fixing the spatial harmonic.

The broad class of real harmonic solutions in Refs. \cite{Matos:1995my,Matos:2000za,Yazadjiev:2000sa} includes this branch. In the spherical sector, these solutions admit an independent scalar flux, as analyzed in Ref. \cite{Porfyriadis:2023qqo}. The regular Gibbons--Maeda black hole is uniquely fixed by the boundary conditions specified below. In the limit \(\al=0\), the fields shown still give \(\phi=-d_0\s/L\), \(f=\mathcal D_e^{-2}\), and the scalar term in \(H\). In this spherical charged scalar-vacuum sector, the family reduces to the Penney solution \cite{Penney:1969xk}.

\subsubsection{Regularity conditions and the Gibbons--Maeda solution}

We now adopt the Schwarzschild harmonic function, keeping a generic \(H\) for the moment in \(k=HK\). In the spherical configuration one has \(H=1\). Along the branch that reaches \(x=1\) with \(\lambda>0\), the denominator approaches a nonzero constant multiple of \(e^{-\lambda\s}\), because \(\s\to-\infty\) in that limit. Hence, from Eq. \eqref{eq:GMharmonic} we find \(f\propto\exp\{2[\lambda+\al(d_0+\al l_1)/L]\s/(1+\al^2)\}\). The scalar field contains a contribution proportional to \([\al\lambda-(d_0+\al l_1)/L]\s\). Requiring the scalar to remain finite imposes \(d_0+\al l_1=\al\lambda L\). Under this condition, the energy reduces to \(H=\lambda^2\), and thus the spherical positive branch picks
\begin{equation}
     \lambda=1,\qquad d_0=\al(L-l_1),\qquad
     L^2=l_1^2+d_0^2-q_0^2>0.
    \label{eq:GMregularity}
\end{equation}

Regularity now constrains both the geodesic energy and its associated scalar term. Imposing \(H=1\) sets the spatial normalization, whereas \(d_0=\al(L-l_1)\) removes the divergent scalar piece at the horizon. Consequently, the secondary scalar charge is determined by enforcing a boundary condition on the geodesic data that were initially independent.

The redshift therefore vanishes linearly at \(x=1\), while the angular area remains finite. Consequently, the scalar field behavior together with the horizon geometry fix the remaining integration constants. In particular, the scalar charge is no longer independent but is fixed by the mass and the electric charge. Removing \(L\) leads to \((1-\al^2)d_0^2+2\al l_1d_0+\al^2q_0^2=0\). The solution branch that connects smoothly to the Schwarzschild limit is \(d_0=-\al q_0^2/[l_1+ \sqrt{l_1^2-(1-\al^2)q_0^2}]\), and the exterior region is constrained by Eq \eqref{eq:GMregularity}.

With these constants one has \(\mathcal D_e=(Lx+l_1-\al d_0)/(L\sqrt{x^2-1})\). Plugging this into the integrated field expressions yields
\begin{equation}
    \begin{aligned}
     f={}&\frac{L(x-1)}{Lx+l_1-\al d_0}
     \left[\frac{L(x+1)}{Lx+l_1-\al d_0}
     \right]^{(1-\al^2)/(1+\al^2)},\\
     \dd s^2={}&-f\dd t^2+\frac{L^2}{f}\dd x^2\\
     &+\frac{L^2(x^2-1)}{f}
     \left[\frac{\dd y^2}{1-y^2}+(1-y^2)\dd\varphi^2\right],\\
     \mathcal A={}&-\frac{q_0}{Lx+l_1-\al d_0}\dd t,\\
     \phi={}&\frac{\al}{1+\al^2}
     \ln\frac{L(x+1)}{Lx+l_1-\al d_0}.
    \end{aligned}
    \label{eq:GMsph}
\end{equation}
Expanding these results at large \(x\) recovers the full set of charges \(l_1,q_0,d_0\). This serves as a straightforward consistency check within the coordinate system employed to build the solution.

Using only \(Lx=r-l_1\), the final version is
\begin{equation}
    \begin{aligned}
     f={}&\frac{r-l_1-L}{r-\al d_0}
     \left(\frac{r-l_1+L}{r-\al d_0}\right)^{(1-\al^2)/(1+\al^2)},\\
     \dd s^2={}&-f\dd t^2+f^{-1}\dd r^2+
     \frac{\Delta}{f}\dd\Omega_2^2,\\
     \mathcal A={}&-\frac{q_0}{r-\al d_0}\dd t,\qquad
     \phi=\frac{\al}{1+\al^2}
     \ln\frac{r-l_1+L}{r-\al d_0}.
    \end{aligned}
    \label{eq:GMradial}
\end{equation}
This is the Gibbons--Maeda solution \cite{Gibbons:1987ps}. In the mass-centered coordinates adopted here, the standard radial combination appearing in their formulas is \(r-\al d_0\). This offset is shown explicitly, most notably in the electric potential. The area of a sphere can be obtained directly from \(4\pi\Delta/f\), with no need to define an additional radial coordinate.

On the outer horizon at \(r=l_1+L\), the area becomes \(4\pi(l_1+L-\al d_0)^{2/(1+\al^2)} (2L)^{2\al^2/(1+\al^2)}\). The scalar field takes the finite value \(\al\ln[2L/(l_1+L-\al d_0)]/(1+\al^2)\). Moreover, one-half of the radial derivative of \(f\) is \([2(l_1+L-\al d_0)]^{-1} [2L/(l_1+L-\al d_0)]^{(1-\al^2)/(1+\al^2)}\), which determines the surface gravity. For the nonextremal exterior region with \(L>0\), all of these quantities stay finite. By contrast, at the inner boundary \(x=-1\) the scalar diverges logarithmically whenever the electric dilaton charge is nonzero, corresponding to the familiar inner scalar singularity in this class of solutions.

For \(\al=1\), Eq \eqref{eq:GMregularity} implies \(d_0=L-l_1=-q_0^2/(2l_1)\). In this case, Eq \eqref{eq:GMradial} reduces to \(f=\Delta/(r^2-d_0^2)\), \(\mathcal A=-q_0\dd t/(r-d_0)\), and \(\phi=\tfrac12\ln[(r+d_0)/(r-d_0)]\), which corresponds to the electric GHS solution \cite{Garfinkle:1990qj}. If one sets \(q_0=0\) on the branch that connects to Schwarzschild, then \(d_0\) must vanish and \(L=l_1\), yielding Eq \eqref{eq:Schwarzschild} with both matter fields turned off.

\subsubsection{The magnetic dilatonic solution}

The magnetic branch is obtained directly from the field equations via the map \((a,b,c)\mapsto(a,-i b,-c)\). The first equation remains the same form since \(|-ib|^2=|b|^2\). For the second equation, reversing the sign of \(c\) while shifting the phase of \(b\) produces the necessary conjugate contribution. Moreover, \((-ib)^2+\overline{(-ib)}^{,2}=-(b^2+\bar b^2)\), so the scalar equation is inverted. Hence, exchanging electric and magnetic fields causes the scalar to switch sign, for a detailed explanation see \cite{Bixano:2026duf}

For an arbitrary harmonic function, we take the electric solution \eqref{eq:GMharmonic} with \(q_0=p_0\) and an electric scalar charge equal to \(-d_0\). The metric coincides with the magnetic metric, whereas the electromagnetic sector is given by \(\mathcal A=-p_0\schdual\dd\varphi/L\) together with the electric scalar with an overall minus sign. In particular, the denominator reads \(\mathcal D_m=\cosh(\lambda\s) -(l_1+\al d_0)\sinh(\lambda\s)/(\lambda L)\), with \(\lambda^2=[(l_1+\al d_0)^2-(1+\al^2)p_0^2]/L^2\). The complete fields for an arbitrary harmonic are
\begin{equation}
    \begin{aligned}
     f={}&\exp\!\left[
     \frac{2\al(-d_0+\al l_1)}{(1+\al^2)L}\s\right]
     \mathcal D_m^{-2/(1+\al^2)},\\
     \dd s^2={}&-f\dd t^2+
     f^{-1}\left[e^{2HK[\s]}(\dd\rho^2+\dd z^2)
     +\rho^2\dd\varphi^2\right],\\
     \mathcal A={}&-\frac{p_0}{L}\schdual\dd\varphi,\\
     \phi={}&\frac{\al}{1+\al^2}\ln\mathcal D_m
     +\frac{-d_0+\al l_1}{(1+\al^2)L}\s,\\
     H={}&\frac{l_1^2+d_0^2-p_0^2}{L^2}.
    \end{aligned}
    \label{eq:GMmagharmonic}
\end{equation}
The scalar derivative at infinity remains \(-d_0/L\), indicating that \(d_0\) represents the magnetic solution’s intrinsic physical scalar charge.

For the regular spheroidal case, the same horizon calculation gives \(d_0=\al(l_1-L)\), \(L^2=l_1^2+d_0^2-p_0^2>0\), and \(H=1\). In the final coordinates,
\begin{equation}
    \begin{aligned}
     f={}&\frac{r-l_1-L}{r+\al d_0}
     \left(\frac{r-l_1+L}{r+\al d_0}\right)^{(1-\al^2)/(1+\al^2)},\\
     \dd s^2={}&-f\dd t^2+f^{-1}\dd r^2+
     \frac{\Delta}{f}\dd\Omega_2^2,\\
     \mathcal A={}&-p_0\cos\theta\dd\varphi,\qquad
     \phi=\frac{\al}{1+\al^2}
     \ln\frac{r+\al d_0}{r-l_1+L}.
    \end{aligned}
    \label{eq:GMmagnetic}
\end{equation}
The magnetic flux equals \(4\pi p_0\), and the area follows from the same angular metric coefficient. A reversed scalar sign separates the electric and magnetic sectors under continuous coupling. However, to construct a dyon with a nontrivial scalar field, the real and imaginary components of \(b\) must be integrated simultaneously. When \(a\) is static, the field equations yield \([\Imm b/\Ree b]'=-2c\,\Imm b/\Ree b\) wherever \(\Ree b\ne0\). This makes clear how the scalar modifies the relative radial scaling of the two components. The two fully integrable dyonic examples presented next illustrate this behavior.

\subsection{Static dyons at \texorpdfstring{\(\al=1\)}{alpha=1}}

\subsubsection{Integration of the electric and magnetic parts}

With \(\al=1\), and choosing \(a(\s)\) and \(c(\s)\) to be real and maintaining both the real and imaginary components of \(b(\s)\). Adding and subtracting the first and third field equations yields \((a+c)'=2(\Ree b)^2\) and \((a-c)'=2(\Imm b)^2\). Meanwhile, the electromagnetic equation separates into \((\Ree b)'=(a+c)\Ree b\) and \((\Imm b)'=(a-c)\Imm b\). Thus, these two combinations satisfy decoupled Riccati integrations, with integration constants fixed by the physical charges.
\begin{equation}
     h_e^2=\frac{(l_1-d_0)^2-2q_0^2}{L^2},\qquad
     h_m^2=\frac{(l_1+d_0)^2-2p_0^2}{L^2}.
    \label{eq:stringconstants}
\end{equation}
Using the same normalized integration as above, and defining the two denominators that occur in the physical fields,
\begin{equation}
    \begin{aligned}
     \mathcal D_e&=\cosh(h_e\s)-\frac{l_1-d_0}{h_eL}\sinh(h_e\s),\\
     \mathcal D_m&=\cosh(h_m\s)-\frac{l_1+d_0}{h_mL}\sinh(h_m\s).
    \end{aligned}
    \label{eq:stringdenominators}
\end{equation}
The linear and real trigonometric cases are relevant whenever one of the constants in \eqref{eq:stringconstants} is zero or takes a negative value. The resulting integrated coefficients are \(a=-\tfrac12(\ln\mathcal D_e\mathcal D_m)'\), \(c=-\tfrac12(\ln[\mathcal D_e/\mathcal D_m])'\), and \(b=-q_0/(L\mathcal D_e)+ip_0/(L\mathcal D_m)\). With these forms, each electromagnetic equation is satisfied independently, and taking their first integrals yields \(H=(h_e^2+h_m^2)/2=(l_1^2+d_0^2-q_0^2-p_0^2)/L^2\).

Integrating the reconstruction equations gives the complete fields for the arbitrary harmonic:
\begin{equation}
    \begin{aligned}
     f={}&\frac{1}{\mathcal D_e\mathcal D_m},\\
     \dd s^2={}&-f\dd t^2+
     f^{-1}\left[e^{2HK[\s]}(\dd\rho^2+\dd z^2)
     +\rho^2\dd\varphi^2\right],\\
     \mathcal A={}&\frac{q_0\sinh(h_e\s)}{h_eL\mathcal D_e}\dd t
     -\frac{p_0}{L}\schdual\dd\varphi,\\
     \phi={}&\frac12\ln\frac{\mathcal D_m}{\mathcal D_e}.
    \end{aligned}
    \label{eq:stringharmonic}
\end{equation}
For instance, \(\sqrt f/\kappa=1/\mathcal D_e\), meaning the electric reconstruction coincides exactly with the primitive already employed in the pure-charge branch. In the same way, \(\kappa\sqrt f=1/\mathcal D_m\) yields the magnetic potential upon integration. This clarifies why two distinct denominators show up: they encode the different scalar weights for the two Maxwell components. Nevertheless, both depend on the same \(\s\).

\subsubsection{The spherical dyon and its scalar-charge relation}

Using Eq \eqref{eq:Schharmonic} with \(H=1\), the regular positive branch implies \(h_e,h_m>0\). In the vicinity of \(x=1\), the redshift scales as \(e^{(h_e+h_m)\s}\), while the scalar field has leading behavior \((h_e-h_m)\s/2\). Hence, for the scalar to remain finite one must have \(h_e=h_m\). Imposing the energy condition then sets \(h_e=h_m=1\), so the redshift vanishes linearly. Taking the difference of the two relations in \eqref{eq:stringconstants} at these parameter values gives
\begin{equation}
     2l_1d_0=p_0^2-q_0^2,\qquad
     L^2=l_1^2+d_0^2-q_0^2-p_0^2>0.
    \label{eq:stringcharges}
\end{equation}
Therefore, for a regular black hole the electric and magnetic charges fix the scalar charge. Nevertheless, the general local solution \eqref{eq:stringharmonic} continues to allow independent scalar constant.

The energy constraint determines the total \(H=(h_e^2+h_m^2)/2\), while requiring the scalar to be regular fixes the relative values by enforcing \(h_e=h_m\). On the positive branch, combining these conditions yields \(h_e=h_m=1\). The scalar–charge relation then captures the requirement that the electric and magnetic parts consistently correspond to the same potential-space geodesic at the horizon.

The denominators now become \(\mathcal D_e=(Lx+l_1-d_0)/(L\sqrt{x^2-1})\) and \(\mathcal D_m=(Lx+l_1+d_0)/(L\sqrt{x^2-1})\). Consequently, \(f=L^2(x^2-1)/[(Lx+l_1)^2-d_0^2]\), \(A_t=-q_0/(Lx+l_1-d_0)\), and \(\phi=\tfrac12\ln[(Lx+l_1+d_0)/(Lx+l_1-d_0)]\). Using the prescribed radial coordinate gives the three fields
\begin{equation}
    \begin{aligned}
     f={}&\frac{\Delta}{r^2-d_0^2},\\
     \dd s^2={}&-f\dd t^2+f^{-1}\dd r^2+
     (r^2-d_0^2)\dd\Omega_2^2,\\
     \mathcal A={}&-\frac{q_0}{r-d_0}\dd t
     -p_0\cos\theta\dd\varphi,\qquad
     \phi=\frac12\ln\frac{r+d_0}{r-d_0}.
    \end{aligned}
    \label{eq:stringdyon}
\end{equation}
This corresponds to the standard spherical dyonic dilaton solution at \(\al=1\) \cite{Gibbons:1987ps,SJPoletti_1995}. The exterior region satisfies \(r>l_1+L>|d_0|\). The horizon has area \(4\pi[(l_1+L)^2-d_0^2]\), with surface gravity \(L/[(l_1+L)^2-d_0^2]\). On this branch, both the scalar field and the electromagnetic sector remain finite when written in a regular horizon gauge. The sign of \(d_0\) changes sign under interchange of the two squared charges. In particular, when \(q_0^2=p_0^2\) one finds \(d_0=0\) for \(l_1>0\), and the fields reduce to the balanced RN solution. If one instead sets one of the charges to zero, the solution becomes the corresponding purely electric or purely magnetic GHS case.

\subsection{The Kaluza--Klein dyon at \texorpdfstring{\(\al^2=3\)}{alpha squared=3}}

\subsubsection{Polynomial integration and the physical charges}

We then consider the standard scalar with \(\al=\sqrt3\), with \(a(\s)\) and \(c(\s)\) real and \(b(\s)\) complex. This choice of coupling yields the static dyonic Toda branch of Kaluza--Klein gravity \cite{Rasheed:1995zv,SJPoletti_1995}. A convenient exact integration can be carried out using two quadratic polynomials in \(e^{2h\s}\):
\begin{equation}
    \begin{aligned}
     \mathcal P_e(\s)&=1-2\delta_e e^{2h\s}
     +\delta_e\delta_m e^{4h\s},\\
     \mathcal P_m(\s)&=1-2\delta_m e^{2h\s}
     +\delta_e\delta_m e^{4h\s}.
    \end{aligned}
    \label{eq:KKpolynomials}
\end{equation}
In this configuration, \(h\) sets the harmonic contribution to the energy, while the two parameters \(\delta_e,\delta_m\) specify the electric and magnetic amplitudes. Their physical values will be set later. We consider a real interval connected to \(\s=0\) where both polynomials remain positive. The normalized redshift and scalar factor are
\begin{equation}
     f=e^{2h\s}\sqrt{\frac{\mathcal P_e(0)\mathcal P_m(0)}
     {\mathcal P_e(\s)\mathcal P_m(\s)}},\qquad
     \kappa=\left[
     \frac{\mathcal P_e(\s)\mathcal P_m(0)}
     {\mathcal P_m(\s)\mathcal P_e(0)}\right]^{3/4}.
    \label{eq:KKpotentials}
\end{equation}
Their logarithmic derivatives determine \(a=h-\tfrac14(\ln[\mathcal P_e\mathcal P_m])'\) and \(c=-\tfrac34(\ln[\mathcal P_e/\mathcal P_m])'\). The real and imaginary electromagnetic components are
\begin{equation}
    \begin{aligned}
     \Ree b={}&-\frac{q_0}{L}e^{h\s}
     \frac{\mathcal P_e(0)}{\mathcal P_e(\s)}
     \sqrt{\frac{\mathcal P_m(\s)}{\mathcal P_m(0)}},\\
     \Imm b={}&\frac{p_0}{L}e^{h\s}
     \frac{\mathcal P_m(0)}{\mathcal P_m(\s)}
     \sqrt{\frac{\mathcal P_e(\s)}{\mathcal P_e(0)}}.
    \end{aligned}
    \label{eq:KKb}
\end{equation}
The charge magnitudes required by the field equations are
\begin{equation}
     \frac{q_0^2}{L^2}=
     \frac{2h^2\delta_e\mathcal P_m(0)}{\mathcal P_e(0)^2},\qquad
     \frac{p_0^2}{L^2}=
     \frac{2h^2\delta_m\mathcal P_e(0)}{\mathcal P_m(0)^2}.
    \label{eq:KKemcharges}
\end{equation}
For a direct check, differentiating the two polynomial ratios gives
\begin{equation}
    \begin{aligned}
     \left[\frac{e^{2h\s}(1-\delta_m e^{2h\s})}
     {\mathcal P_e(\s)}\right]'&=
     \frac{2h e^{2h\s}\mathcal P_m(\s)}{\mathcal P_e(\s)^2},\\
     \left[\frac{e^{2h\s}(1-\delta_e e^{2h\s})}
     {\mathcal P_m(\s)}\right]'&=
     \frac{2h e^{2h\s}\mathcal P_e(\s)}{\mathcal P_m(\s)^2}.
    \end{aligned}
    \label{eq:KKidentities}
\end{equation}
Combined with the logarithmic derivatives in Eq \eqref{eq:KKb}, these identities yield \(a'=|b|^2\), \(c'=3[(\Ree b)^2-(\Imm b)^2]\), as well as the complete complex equation for \(b\). The energy remains conserved and is given by \(H=h^2\).

The mass and scalar charge are:
\begin{equation}
    \begin{aligned}
     \frac{l_1}{L}={}&h+
     \frac{h\delta_e(1-\delta_m)}{\mathcal P_e(0)}+
     \frac{h\delta_m(1-\delta_e)}{\mathcal P_m(0)},\\
     \frac{d_0}{L}={}&-
     \frac{\sqrt3h\delta_e(1-\delta_m)}{\mathcal P_e(0)}+
     \frac{\sqrt3h\delta_m(1-\delta_e)}{\mathcal P_m(0)}.
    \end{aligned}
    \label{eq:KKmassscalar}
\end{equation}
In particular, one has \(h^2L^2=l_1^2+d_0^2-q_0^2-p_0^2\). From Eqs. \eqref{eq:KKemcharges} and \eqref{eq:KKmassscalar} it also follows that \(2d_0(d_0^2-3l_1^2) =3q_0^2(d_0+\sqrt3l_1)+3p_0^2(d_0-\sqrt3l_1)\). This polynomial identity encodes the scalar-charge constraint for this exact dyonic branch, without requiring division by an individual charge or by any potentially vanishing combination.

\subsubsection{Reconstruction and the spheroidal black hole}

For \(q_0\ne0\), it is 
\begin{equation}
     A_t=\frac{hL}{q_0}\left[
     \frac{1-\delta_e e^{2h\s}}{\mathcal P_e(\s)}
     -\frac{1-\delta_e}{\mathcal P_e(0)}\right].
    \label{eq:KKAt}
\end{equation}
Its derivative equals \(q_0f/(L\kappa^2)\), and at \(\s=0\) it evaluates to zero. When the electric charge is zero, the reconstruction equation immediately yields \(A_t=0\). Therefore, this endpoint is preserved without having to divide by a vanishing charge. The most general arbitrary-harmonic configuration is
\begin{equation}
    \begin{aligned}
     \dd s^2={}&-f\dd t^2+
     f^{-1}\left[e^{2h^2K[\s]}(\dd\rho^2+\dd z^2)
     +\rho^2\dd\varphi^2\right],\\
     \mathcal A={}&A_t\dd t-\frac{p_0}{L}\schdual\dd\varphi,\\
     \phi={}&-\frac{\sqrt3}{4}\ln\left[
     \frac{\mathcal P_e(\s)\mathcal P_m(0)}
     {\mathcal P_m(\s)\mathcal P_e(0)}\right],
    \end{aligned}
    \label{eq:KKharmonic}
\end{equation}
where \(f\) and \(A_t\) are given explicitly in Eqs \eqref{eq:KKpotentials} and \eqref{eq:KKAt}. We keep this polynomial branch since it includes the usual static dyonic Kaluza--Klein black hole, for which both charges are nonvanishing.

For the Schwarzschild harmonic, choosing the positive branch \(h>0\) yields finite polynomial limits at \(x=1\) because \(e^{2h\s}\to 0\). Consequently, the scalar stays bounded. The redshift scales like \(e^{2h\s}\), while the angular area density carries the factor \((x-1)^{(h-1)^2/2}\). Taking \(h=1\) implies \(H=1\), producing a simple zero in the redshift and a finite, nonvanishing angular area. In spheroidal coordinates, the fields are
\begin{equation}
    \begin{aligned}
     \dd s^2={}&-f\dd t^2+\frac{L^2}{f}\dd x^2\\
     &+\frac{L^2(x^2-1)}{f}
     \left[\frac{\dd y^2}{1-y^2}+(1-y^2)\dd\varphi^2\right],\\
     \mathcal A={}&A_t\dd t-p_0y\dd\varphi,\qquad
     \phi=-\frac{\sqrt3}{4}\ln\frac{\mathcal P_e(\s)\mathcal P_m(0)}
     {\mathcal P_m(\s)\mathcal P_e(0)},
    \end{aligned}
    \label{eq:KKsph}
\end{equation}
with \(e^{2\s}=(x-1)/(x+1)\). Imposing the last coordinate relation \(Lx=r-l_1\) turns this ratio into \((r-l_1-L)/(r-l_1+L)\), consequently the metric takes the form \(-f\dd t^2+f^{-1}\dd r^2+(\Delta/f)\dd\Omega_2^2\), with the four-potential and scalar from Eq \eqref{eq:KKsph} evaluated using this same ratio. Therefore, all resulting functions can be written explicitly as rational expressions and real powers of the given radial combination.

The horizon lies at \(r=l_1+L\). Its area equals \(16\pi L^2/\sqrt{\mathcal P_e(0)\mathcal P_m(0)}\), while the scalar on the horizon is \(\sqrt3\,\ln[\mathcal P_e(0)/\mathcal P_m(0)]/4\). From the first polynomial antiderivative and from the metric one likewise finds the surface gravity to be \(\sqrt{\mathcal P_e(0)\mathcal P_m(0)}/(4L)\). The physically relevant exterior region is picked out by requiring both polynomials to be positive for \(\s\le0\). Switching off either charge yields the associated Gibbons--Maeda solution with \(\al^2=3\). If the polynomial parameters are chosen equal, the squared charges coincide and \(d_0=0\), which brings one back to the balanced RN case. Leaving \(h\ne1\) preserves the exact harmonic deformation in \eqref{eq:KKharmonic} and the corresponding Weyl exponent \(h^2\).

Within this polynomial branch, \(h\) sets the normalization of the geodesic energy, whereas \(\delta_e,\delta_m\) encode the electromagnetic and scalar parameters, constrained by the charge relation shown. Imposing the horizon condition enforces \(h=1\), thereby fixing the energy normalization while leaving both electromagnetic charges free. The scalar-charge relation, however, already holds throughout the polynomial branch even before applying the horizon condition.

\section{Wormholes in the harmonic branches}\label{sec:wormholes}

An analogous harmonic reconstruction can be used to characterize wormhole spacetimes as well. In the scalar-vacuum regime, the pertinent nonsingular configurations are the Ellis--Bronnikov wormholes \cite{Ellis:1973yv,Bronnikov:1973fh}. We keep the normalization and physical parameters defined earlier, and employ the oblate harmonic introduced in Sec. \ref{sec:zero-negative-energy}. The remaining freedom is the sign choice for the scalar kinetic term. Here we modify Eq \eqref{eq:action} by substituting \(2(\nabla\phi)^2\) with \(2\epsilon_0(\nabla\phi)^2\), where \(\epsilon_0=+1\) corresponds to the scalar sign adopted above and \(\epsilon_0=-1\) denotes the phantom sign.

\subsection{Scalar-vacuum integration and oblate regularity}

With this sign included, the first two equations in \eqref{eq:abc} retain their form, while the scalar equation and the conserved energy become
\begin{equation}
    \begin{aligned}
    c'&=\frac{\al^2}{\epsilon_0}\Ree(b^2),\\
    H&=|a|^2-|b|^2+\frac{\epsilon_0c^2}{\al^2}.
    \end{aligned}
    \label{eq:WHenergy}
\end{equation}
When \(\al=0\), the scalar term is computed explicitly as \(\epsilon_0(\phi')^2\). This keeps the free scalar as an independent degree of freedom, without relying on any quotient that would involve the zero coupling.

We enforce staticity by choosing \(a(\s)\) to be real and eliminate the electromagnetic sector by setting \(b(\s)=0\). The resulting equations imply \(a'=c'=0\). Integrating these, and using the asymptotic data in Eq \eqref{eq:charges}, we obtain \(a=l_1/L\), \(c=-\al d_0/L\), and \(H=(l_1^2+\epsilon_0d_0^2)/L^2\). Both the NUT and Maxwell charges are zero, whereas \(l_1\) and \(d_0\) remain independent until regularity conditions are applied. The fields take the form given in Eq \eqref{eq:scalarharmonic} with this choice of \(H\), reconstructing them yields \(f=e^{2l_1\s/L}\), \(\mathcal A=0\), \(\phi=-d_0\s/L\), and \(k=HK[\s]\). When \(\al=0\), the same \(\phi\) follows directly from the free scalar equation. Because the Maxwell field is absent, the solution is valid for any continuous value of \(\al\).

We now adopt the oblate harmonic introduced in Sec. \ref{sec:zero-negative-energy}, \(\s=\arctan x-\pi/2\), and take \(-\infty<x<\infty\), with \(-1<y<1\) and \(\varphi\) having period \(2\pi\). Its asymptotic normalization as \(x\to+\infty\) is \(\s=-x^{-1}+O(x^{-3})\), ensuring that the parameters \(l_1,d_0\) keep their interpretation as charges at that end. The harmonic condition is obtained from \((x^2+1)\s_{,x}=1\). Applying the oblate coordinate transformation and the corresponding Weyl quadrature given there, one finds that the complete solution
\begin{equation}
    \begin{aligned}
    \dd s^2={}&-f\dd t^2+\frac{L^2}{f}
    \Bigg\{\left(\frac{x^2+y^2}{x^2+1}\right)^{H+1}\\
    &\times\left[\dd x^2+\frac{x^2+1}{1-y^2}\dd y^2\right]\\
    &+(x^2+1)(1-y^2)\dd\varphi^2\Bigg\},\\
    f={}&e^{2l_1\s/L},\qquad
    \mathcal A=0,\qquad \phi=-\frac{d_0}{L}\s.
    \end{aligned}
    \label{eq:WHoblate}
\end{equation}
The exponent \(H+1\) incorporates both the term \(H\) coming from the potential-space geodesic and the effect of the oblate coordinate mapping. This total exponent is the one that needs to be considered to ensure a smooth continuation.

The transformation to \((\rho,z)\) becomes singular at \(x=y=0\), which corresponds to the ring \(\rho=L, z=0\). To probe the behavior of the metric near this ring, introduce local coordinates \(x=u\cos\beta\), \(y=u\sin\beta\), where on the full oblate extension \(\beta\) has period \(2\pi\). Holding \(t,\varphi\) fixed, the leading-order transverse line element is \(L^2 e^{\pi l_1/L} u^{2(H+1)}(\dd u^2 + u^2 \dd\beta^2)\). When \(H>-2\), the ratio of the circumference to the proper distance from \(u=0\) approaches \(2\pi(H+2)\). Smoothness of the interior circle demands this ratio equal \(2\pi\), which sets \(H=-1\). If \(H\le -2\), the ring is at infinite meridional distance and therefore does not yield the smooth extension considered here. Hence, in this scalar-vacuum class and for this choice of harmonic normalization, regularity implies
\begin{equation}
    H=-1,\qquad \epsilon_0=-1,\qquad
    L^2=d_0^2-l_1^2>0.
    \label{eq:WHregularity}
\end{equation}
With the usual scalar sign, one finds \(H\ge0\), so this requirement cannot be met. A phantom scalar instead provides the needed negative term to offset the oblate meridional factor. The other fields must be examined as well: the superextremal electrovacuum case in Sec. \ref{sec:zero-negative-energy} exhibits the same normalized energy, but has a singular Maxwell field and a singular curvature at the boundary.

\subsection{The massive and ultrastatic Ellis--Bronnikov solutions}

Inserting Eq \eqref{eq:WHregularity} into \eqref{eq:WHoblate} removes its dependence on \(x^2+y^2\). The reconstructed solution is
\begin{equation}
    \begin{aligned}
    \dd s^2={}&-f\dd t^2+\frac{L^2}{f}\dd x^2\\
    &+\frac{L^2(x^2+1)}{f}
    \left[\frac{\dd y^2}{1-y^2}+(1-y^2)\dd\varphi^2\right],\\
    f={}&e^{2l_1\s/L},\qquad
    \mathcal A=0,\qquad \phi=-\frac{d_0}{L}\s,\\
    \s={}&\arctan x-\frac{\pi}{2},\qquad
    d_0^2=l_1^2+L^2.
    \end{aligned}
    \label{eq:EBspheroidal}
\end{equation}
This corresponds to the massive Ellis--Bronnikov family \cite{Ellis:1973yv,Bronnikov:1973fh,Gibbons:2016bok}. The connection between the mass and the scalar charge can be obtained directly from the geodesic energy and the regularity condition. The parameter \(L>0\) sets the geometric scale, whereas the two possible signs of \(d_0\) produce opposite scalar configurations while leaving the metric unchanged.

The field equations may be verified directly, without the need to introduce any extra radial coordinate. The scalar equation simplifies to \(\partial_x[(x^2+1)\phi_{,x}]=0\), because \((x^2+1)\phi_{,x}=-d_0/L\). The only nonvanishing Ricci component is \(R_{xx}=-2(L^2+l_1^2)/[L^2(x^2+1)^2]\). This matches \(-2(\phi_{,x})^2\) exactly provided that \(d_0^2=l_1^2+L^2\), in agreement with the phantom Einstein equation. The Ricci scalar is \(R=-2d_0^2f/[L^4(x^2+1)^2]\), and it remains finite everywhere. More broadly, using the angular variable \(y=\cos\theta\), all metric coefficients are smooth for finite \(x\), aside from the standard removable singularities of polar coordinates on the symmetry axis. The lapse stays positive and is bounded away from zero. Hence the metric admits a smooth extension across \(x=0\), including the circle where the Weyl map becomes degenerate.

As \(x\to+\infty\), we have \(f\to1\) and \(\phi\to0\), while as \(x\to-\infty\), \(f\to e^{-2\pi l_1/L}\) and \(\phi\to\pi d_0/L\). Both limits correspond to asymptotically flat regions. At the second end, the constant asymptotic prefactors in the metric can be eliminated by rescaling the time coordinate by \(e^{-\pi l_1/L}\) and the spatial scales by \(e^{\pi l_1/L}\).
The mass evaluated at the first end equals \(l_1\), in contrast, using the second end’s own asymptotic normalization, the mass becomes \(-l_1e^{\pi l_1/L}\). The associated scalar charge at that end, defined with respect to its asymptotic scalar value, is \(-d_0e^{\pi l_1/L}\). Therefore, despite regularity at both ends, a massive configuration displays an intrinsic asymmetry between them.

At fixed \(x\), the surface area of the sphere is \(\operatorname{Area}(x)=4\pi L^2(x^2+1)/f\). Taking its logarithmic derivative gives \(\partial_x\ln\operatorname{Area}=2(x-l_1/L)/(x^2+1)\). There is a single minimum located at \(x=l_1/L\), and at this point \(\partial_x^2\ln\operatorname{Area}=2/[1+(l_1/L)^2]>0\). This minimal sphere defines the wormhole throat. When the mass is nonzero, the throat shifts away from \(x=0\), which is the disk appearing in the oblate Weyl transformation. The phantom scalar’s negative kinetic term breaks the null energy condition, providing the support for this minimum. Radial null geodesics obey \(\dd t/\dd x=\pm L/f\), remaining finite as they pass through the throat. The geometry has two asymptotically flat ends at infinite radial proper distance, with no horizon in between.

At this point, we apply Eq \eqref{eq:BL} together with \(Lx=r-l_1\) and \(y=\cos\theta\). The resulting radial expressions for all three fields are
\begin{equation}
    \begin{aligned}
    f={}&\exp\left\{\frac{2l_1}{L}
    \left[\arctan\frac{r-l_1}{L}-\frac{\pi}{2}\right]\right\},\\
    \dd s^2={}&-f\dd t^2+f^{-1}\dd r^2\\
    &+f^{-1}\left[(r-l_1)^2+L^2\right]\dd\Omega_2^2,\\
    \mathcal A={}&0,\qquad
    \phi=-\frac{d_0}{L}
    \left[\arctan\frac{r-l_1}{L}-\frac{\pi}{2}\right].
    \end{aligned}
    \label{eq:EBradial}
\end{equation}
The coordinate \(r\) spans the entire real line, with the throat located at \(r=2l_1\). This position is determined by the condition of minimal area, and \(r\) is the mass-centered coordinate as defined in Eq \eqref{eq:BL}.

\paragraph{Ultrastatic solution.}
Setting \(l_1=0\) gives \(d_0=\pm L\) and \(f=1\). The solution reduce to
\begin{equation}
    \begin{aligned}
    \dd s^2={}&-\dd t^2+L^2\dd x^2\\
    &+L^2(x^2+1)
    \left[\frac{\dd y^2}{1-y^2}+(1-y^2)\dd\varphi^2\right],\\
    \mathcal A={}&0,\qquad
    \phi=-\frac{d_0}{L}
    \left(\arctan x-\frac{\pi}{2}\right),
    \qquad d_0^2=L^2.
    \end{aligned}
    \label{eq:EBultrastatic}
\end{equation}
This describes the ultrastatic Ellis–Bronnikov wormhole. The throat lies at \(x=0\) and has area \(4\pi L^2\), while the redshift remains uniform across the entire spacetime. The asymptotic masses on both sides are zero, and the throat is maintained by a nonvanishing phantom scalar field.

\subsection{The vacuum ring and the role of the global extension}

The same oblate chart also makes the vacuum comparison explicit. Set \(l_1=d_0=0\), so that \(a=b=c=0\), \(H=0\), and all potentials are constant. Equation
\eqref{eq:WHoblate} becomes
\begin{equation}
    \begin{aligned}
    \dd s^2={}&-\dd t^2+L^2(x^2+y^2)
    \left[\frac{\dd x^2}{x^2+1}
    +\frac{\dd y^2}{1-y^2}\right]\\
    &+L^2(x^2+1)(1-y^2)\dd\varphi^2,\\
    \mathcal A={}&0,\qquad \phi=0.
    \end{aligned}
    \label{eq:vacuumring}
\end{equation}
Under the oblate transformation, this line element is locally brought to the form \(-\dd t^2+\dd\rho^2+\dd z^2+\rho^2\dd\varphi^2\). On the standard single-sheet domain, it represents Minkowski spacetime. If, instead, one retains both oblate sheets and extends them through the disk, the result is the vacuum ring wormhole \cite{Gibbons:2016bok}.

The conditions at the ring are determined by the transverse test carried out earlier. When \(H=0\), the ratio of the circumference to the proper radial distance at the ring equals \(4\pi\). This implies a conical excess of \(2\pi\) and a distributional curvature term corresponding to negative tension localized on the ring. Away from this source the spacetime is vacuum and flat. As a result, two asymptotic regions are present together with a ring singularity. A similar regularization using a phantom scalar field is also found in stationary ring geometries \cite{Volkov:2026rpu}.

In potential space, the vacuum state appears as a constant trajectory with vanishing tangent. With \(H=0\), it thus falls into the degenerate category, while a nonconstant null geodesic corresponds to distinct choices of initial data. By contrast, the regular Ellis--Bronnikov branch has a nonzero scalar tangent and satisfies \(H=-1\). Via \(k=HK[\s]\), this negative scalar term exactly cancels the ring defect in the static metric. Along with the prolate black-hole cases, this illustrates within the same framework, how the conserved energy, the chosen harmonic, and the boundary behavior of the reconstructed fields jointly determine the regularity requirements.

\subsection{Other harmonics and stationary wormholes}\label{sec:other-WH-harmonics}

The stationary configurations reported in Refs.\ \cite{Bixano:2025jwm,Bixano:2025bio} provide an alternative realization of the same harmonic system. In these solutions, the electric, magnetic, twist, and scalar sectors are all turned on simultaneously. Adopting the gauge choices for the potentials and the electromagnetic normalization used in this work, one can take a representative geodesic with \(a=i/2\), \(b=(1+i)/2\), and \(c=-1\). Plugging these values into Eqs.\ \eqref{eq:abc} and \eqref{eq:WHenergy} yields \(a'=b'=c'=0\) and \(H=\epsilon_0/\al^2-1/4\), with \(\al\ne0\). Specifically, the relation \(|b|^2=2(\Imm a)^2\) ensures that the electromagnetic and twist contributions cancel in the first equation, whereas \(\Ree(b^2)=0\) maintains a constant scalar tangent. This singles out an exact branch with complex \(a(\s),b(\s)\) and real \(c(\s)\), without revisiting the associated spacetime reconstruction. It represents one fully active branch: its integration constants obey these constraints but do not span all possible initial data of the full system.

The key difference from the earlier monopole cases is the choice of spatial harmonic. In those studies, the oblate harmonics are taken proportional to \(x/(x^2+y^2)\), \(y/(x^2+y^2)\), or a linear combination of the two, and they also treat a distinct harmonic proportional to \(xy\). While the geodesic equations are unchanged, \(K[\s]\) and \(\schdual\) are modified. As a result, the properties of the ring, the angular dependence, and the asymptotic interpretation are altered too. The charges must therefore be read off from the asymptotic expansion of the selected harmonic: Eq. \eqref{eq:charges} assumes the monopole normalization \(\s=-x^{-1}+O(x^{-2})\). If the harmonic instead begins with a dipole term, the same monopole charges cannot be inferred just from \(a(0),b(0),c(0)\).

Ref.\ \cite{Bixano:2025qxp} studies the ring’s causal structure and its connection to a throat, while Ref.\ \cite{Bixano:2025kfm} reviews wormhole cosmic censorship. In these scenarios a curvature singularity remains, and the question is whether causal geodesics can reach it. By contrast, a regular throat with a nonsingular continuation requires the metric and matter fields to be well behaved and smoothly extend across the region. Stating both problems explicitly allows a unified framework for singular stationary branches and smooth phantom-supported wormholes, with the appropriate global conditions in each case.

Ref.\ \cite{Sarmiento-Alvarado:2026efk} studies related vacuum setups, linking the asymptotically flat branch to the Gibbons--Volkov geometry and completing it via a cut-and-paste procedure. Ref.\ \cite{Sarmiento-Alvarado:2026prq} gives a five-dimensional Ricci-flat example showing that the choice of completion can matter, though its field content lies beyond the four-dimensional EMD framework used here. Any junction-based construction must also specify junction conditions and the associated surface stress tensor, which are additional inputs beyond a smooth bulk reconstruction.


\section{Conclusions}   %

The unified Ernst framework and the Hamiltonian reconstruction offer a shared geometric perspective on the solutions analyzed here \cite{Bixano:2026xum,Bixano:2026duf}. Performing the integration selects a geodesic in potential space, with normalized initial conditions that capture the physical charges. The associated conserved energy determines the meridional Weyl quadrature, whereas the selected harmonic specifies how these data are implemented in spacetime. Consequently, the regularity requirements obtained for the separate branches constrain both the geodesic data and the manner in which they are spatially realized.

For the Schwarzschild harmonic, the regular nonextremal horizons studied here enforce \(H=1\), along with branch-specific constraints on the scalar and electromagnetic fields. The Fisher solution clearly shows that, besides fixing the conserved energy, one must also regulate the scalar configuration. In the oblate scalar-vacuum family, instead, smooth extension picks \(H=-1\), which gives \(L^2=d_0^2-l_1^2\) and leads to the Ellis--Bronnikov wormholes. The vacuum ring highlights that the global completion can matter even when all potentials remain constant. This work’s contribution is to make explicit how the \(a,b,c\) integrations relate to these boundary conditions and to the consistent reconstruction of all three physical fields within a single set of conventions.

Conserved quantities in potential space become constraints on spacetime fields once a harmonic is chosen. Mass, NUT, electric/magnetic, and scalar charges fix the initial tangent and its norm, the associated Noether integrals reconstruct the twist and electromagnetic potential. Via \(k=HK[\s]\), this norm also fixes the meridional geometry, so regularity selects compatible geodesic data and spatial realizations rather than allowing an independent metric adjustment after fixing matter. Reconstruction and boundary conditions must be satisfied jointly by \(\dd s^2\), \(\mathcal A\), and \(\phi\). Appendix \ref{app:chargedZV} maps the parameters to the stationary charged Zipoy--Voorhees solution, including its electromagnetic duality freedom.

This is also constructive. By continuous dependence on initial data, each Ellis--Bronnikov solution has an open neighborhood of regular static configurations with both electromagnetic charges nonzero, for any fixed finite real \(\al\). The energy stays \(H=-1\), the meridional cancellation remains exact, and all three fields extend smoothly between two asymptotically flat ends. The result requires neither special integrable couplings nor truncated charge expansions, but it is local in initial data: it does not bound the maximal charges, establish dynamical stability, or replace a global uniqueness theorem.

In this sense, black holes and wormholes are related within the harmonic construction. The scalar-vacuum reduction already includes the Schwarzschild and Ellis–Bronnikov solutions, and the charged equations organize the corresponding matter sectors in the same framework. With the present normalization, a regular nonextremal prolate horizon requires \(H=1\), while a smooth oblate throat requires \(H=-1\). These choices correspond to different geodesic data and boundary conditions. Flipping the scalar sign also changes the theory. Because a potential-space isometry preserves the geodesic norm, identifying the two sectors is not, by itself, a symmetry or a proven continuous transition between regular spacetimes. Their link is their shared reduced equations and reconstruction, with regularity determining the allowed realization.

\section{Acknowledgements}

LB thanks SECIHTI-M\'exico for the doctoral grant.
This work was also partially supported by SECIHTI M\'exico under grants SECIHTI CBF-2025-G-1720 and CBF-2025-G-176.

\section*{Data Availability Statement}

This article presents entirely theoretical research. No datasets were generated or analyzed during the current study.


\begin{appendices}

\section{Comparison with the stationary charged Zipoy--Voorhees solution}\label{app:chargedZV}
The stationary metric introduced by Halilsoy, Mangut, and Hsieh \cite{Halilsoy:2022pzr} enables a direct comparison that goes beyond the spherical Brill case. We provide the parameter dictionary, since matching only a limiting solution would not by itself fix the interpretation of the full deformation. Only in this appendix, we denote their charge parameter by \(p_H\), their NUT parameter by \(n_H\), their length scale by \(m_H\), and their distortion by \(\gamma\). The subscript serves to distinguish these quantities from the physical charges already specified in Eq \eqref{eq:charges}.

Let \(k_H=\sqrt{1+n_H^2}\) and \(\mathcal A_H=k_H(1+p_H^2)-(1-p_H^2)>0\). On a common real exterior, the redshift in their Eqs \eqref{eq:ZVmoments}--\eqref{eq:ZVradial}, after adopting the signature of Eq \eqref{eq:weyl} and normalizing time at infinity, reads
\begin{equation}
    \begin{aligned}
     \mathcal A_H f^{-1}={}&k_H(1+p_H^2)\cosh(2\gamma\s)\\
     &-2p_H\sinh(2\gamma\s)-(1-p_H^2).
    \end{aligned}
    \label{eq:HMHredshift}
\end{equation}
The simplification uses \((k_H\cosh u-\sinh u)(k_H\cosh u+\sinh u) =1+n_H^2\cosh^2u\). The argument \(u=\gamma\s\) is used only in this identity.
It removes the rational denominator in their representation and exposes the same hyperbolic dependence as \(|\mathcal D|^2\) in Eq \eqref{eq:complexEMabc}.

The parameter identification is
\begin{equation}
    \begin{aligned}
     h&=\gamma,\qquad L=m_Hp_H\sqrt{\mathcal A_H/2},\\
     l_1&=\frac{2p_HLh}{\mathcal A_H},\qquad
     l_2=-\frac{n_H(1+p_H^2)Lh}{\mathcal A_H},\\
     q_0^2+p_0^2&=\frac{2(1-p_H^2)L^2h^2}{\mathcal A_H}.
    \end{aligned}
    \label{eq:HMHmap}
\end{equation}
To check this, expand the denominator in the current integral:
\(|\mathcal D|^2=\cosh^2(h\s) - \frac{2l_1}{hL}\cosh(h\s)\sinh(h\s) + (l_1^2+l_2^2)\frac{\sinh^2(h\s)}{h^2L^2}\).
In the energy relation in \eqref{eq:complexEMabc}, the final coefficient is instead written as \(1+(q_0^2+p_0^2)/(h^2L^2)\).
Applying the double-angle formulas then reproduces Eq \eqref{eq:HMHredshift} exactly under the map \eqref{eq:HMHmap}. The twist term is \(2l_2y\), and the meridional exponent satisfies \(\gamma^2=h^2\), therefore the full metrics coincide. This correspondence (dictionary) is only defined on the intersection of the real parameter ranges, with the orientation and sign conventions for lengths chosen coherently. Meanwhile, the initial-value integral remains independently valid on its respective real domains.

One can compare the electromagnetic fields directly, without identifying a magnetic Ernst potential with the azimuthal component of the four-potential. For a given metric within this one-harmonic electrovacuum family, the metric functions obey \(b'=(\Ree a-2i\Imm a)b\). Hence, for any two nonvanishing solutions \(b\), their quotient must be a constant. Requiring that they share the same energy constrains the modulus of this constant to unity, so only a constant Maxwell duality rotation angle remains. Fixing that angle then determines \(q_0\) and \(p_0\) individually. With identical boundary data, Eqs. \eqref{eq:reconstruct} and \eqref{eq:quadratures} uniquely specify \(A_t\), \(\chi\), and \(A_\varphi\) up to a gauge transformation. In both descriptions the scalar field vanishes. This finishes the matching of all three physical fields appearing in Eq. \eqref{eq:complexEMharmonic}.

\end{appendices}

\clearpage
\bibliographystyle{elsarticle-harv} 
\bibliography{Bibliografia}

@article{Matos:2000za,
    author = "Matos, Tonatiuh and Nunez, Dario and Rios, Maribel",
    title = "{Class of Einstein-Maxwell dilatons, an ansatz for new families of rotating solutions}",
    eprint = "gr-qc/0008068",
    archivePrefix = "arXiv",
    doi = "10.1088/0264-9381/17/18/323",
    journal = "Class. Quant. Grav.",
    volume = "17",
    pages = "3917--3934",
    year = "2000"
}

@article{Bixano:2025jwm,
    author = "Bixano, Leonel and Matos, Tonatiuh",
    title = "{Einstein-Maxwell-dilaton wormholes that meet the energy conditions}",
    eprint = "2502.07206",
    archivePrefix = "arXiv",
    primaryClass = "gr-qc",
    reportNumber = "PhysRevD.111.084056",
    doi = "10.1103/PhysRevD.111.084056",
    journal = "Phys. Rev. D",
    volume = "111",
    number = "8",
    pages = "084056",
    year = "2025"
}

@article{Bixano:2025bio,
    author = "Bixano, Leonel and Matos, Tonatiuh",
    title = "{Possibility of the existence of wormholes in nature}",
    eprint = "2505.20167",
    archivePrefix = "arXiv",
    primaryClass = "gr-qc",
    doi = "10.1103/mnkg-8ccb",
    journal = "Phys. Rev. D",
    volume = "113",
    number = "8",
    pages = "084004",
    year = "2026"
}

@article{Ernst:1967wx,
    author = "Ernst, Frederick J.",
    title = "{New formulation of the axially symmetric gravitational field problem}",
    doi = "10.1103/PhysRev.167.1175",
    journal = "Phys. Rev.",
    volume = "167",
    pages = "1175--1179",
    year = "1968"
}

@article{Ernst:1967by,
    author = "Ernst, Frederick J.",
    title = "{New Formulation of the Axially Symmetric Gravitational Field Problem. II}",
    doi = "10.1103/PhysRev.168.1415",
    journal = "Phys. Rev.",
    volume = "168",
    pages = "1415--1417",
    year = "1968"
}

@article{Rasheed:1995zv,
    author = "Rasheed, Dean",
    title = "{The Rotating dyonic black holes of Kaluza-Klein theory}",
    eprint = "hep-th/9505038",
    archivePrefix = "arXiv",
    reportNumber = "DAMTP-R-95-25",
    doi = "10.1016/0550-3213(95)00396-A",
    journal = "Nucl. Phys. B",
    volume = "454",
    pages = "379--401",
    year = "1995"
}

@article{Clement:2015aka,
    author = "Cl{\'e}ment, G{\'e}rard and Gal'tsov, Dmitri and Guenouche, Mourad",
    title = "{NUT wormholes}",
    eprint = "1509.07854",
    archivePrefix = "arXiv",
    primaryClass = "hep-th",
    reportNumber = "LAPTH-047-15",
    doi = "10.1103/PhysRevD.93.024048",
    journal = "Phys. Rev. D",
    volume = "93",
    number = "2",
    pages = "024048",
    year = "2016"
}

@article{Gibbons:1987ps,
    author = "Gibbons, G. W. and Maeda, Kei-ichi",
    title = "{Black Holes and Membranes in Higher Dimensional Theories with Dilaton Fields}",
    reportNumber = "UTAP-48-87, LPTENS-87-10",
    doi = "10.1016/0550-3213(88)90006-5",
    journal = "Nucl. Phys. B",
    volume = "298",
    pages = "741--775",
    year = "1988"
}

@article{Garfinkle:1990qj,
    author = "Garfinkle, David and Horowitz, Gary T. and Strominger, Andrew",
    title = "{Charged black holes in string theory}",
    reportNumber = "UCSB-TH-90-66",
    doi = "10.1103/PhysRevD.43.3140",
    journal = "Phys. Rev. D",
    volume = "43",
    pages = "3140",
    year = "1991",
    note = "[Erratum: Phys.Rev.D 45, 3888 (1992)]"
}

@article{Bixano:2026xum,
    author = "Bixano, Leonel and Matos, Tonatiuh",
    title = "{Unified Ernst potentials formalism for Einstein-Maxwell scalar fields theories}",
    journal = "General Relativity and Gravitation",
    volume = "58",
    number = "10",
    pages = "119",
    year = "2026",
    month = "10",
    doi = "10.1007/s10714-026-03621-3",
    url = "https://doi.org/10.1007/s10714-026-03621-3",
    eprint = "2603.02384",
    archivePrefix = "arXiv",
    primaryClass = "gr-qc"
}

@article{Bixano:2026duf,
    author = "Bixano, Leonel and Matos, Tonatiuh",
    title = "{Potential Space Symmetries in Ernst-like Formulations of Einstein-Maxwell/ModMax-Scalar field Theories}",
    eprint = "2605.17843",
    archivePrefix = "arXiv",
    primaryClass = "gr-qc",
    month = "5",
    year = "2026"
}

@article{Bixano:2025qxp,
    author = "Bixano, Leonel and Matos, Tonatiuh",
    title = "{The space-time structure of an untouchable naked singularity in superstrings theory}",
    eprint = "2508.01820",
    archivePrefix = "arXiv",
    primaryClass = "gr-qc",
    month = "8",
    year = "2025"
}

@article{Bixano:2025kfm,
    author = "Bixano, Leonel and Sarmiento-Alvarado, I. A. and Matos, Tonatiuh",
    title = "{A Brief Review of Wormhole Cosmic Censorship}",
    eprint = "2602.16116",
    archivePrefix = "arXiv",
    primaryClass = "gr-qc",
    doi = "10.3390/axioms14110831",
    journal = "Axioms",
    volume = "14",
    number = "11",
    pages = "831",
    year = "2025"
}

@article{Matos:1995my,
    author = "Matos, Tonatiuh and Nunez, Dario and Quevedo, Hernando",
    title = "{Class of Einstein-Maxwell dilatons}",
    eprint = "gr-qc/9510042",
    archivePrefix = "arXiv",
    doi = "10.1103/PhysRevD.51.R310",
    journal = "Phys. Rev. D",
    volume = "51",
    pages = "R310--R313",
    year = "1995"
}

@article{Bixano:2026ouq,
  author = {Bixano, Leonel and Matos, Tonatiuh},
  title = {{Generalized Einstein-ModMax-ScalarField theories and new exact solutions}},
  eprint = {2603.26073},
  archiveprefix = {arXiv},
  primaryclass = {gr-qc},
  month = {3},
  year = {2026},
}

@article{Voorhees:1970ywo,
    author = "Voorhees, B. H.",
    title = "{Static axially symmetric gravitational fields}",
    doi = "10.1103/PhysRevD.2.2119",
    journal = "Phys. Rev. D",
    volume = "2",
    pages = "2119--2122",
    year = "1970"
}

@article{Quevedo:2012ttw,
    author = "Quevedo, Hernando and Toktarbay, Saken and Yerlan, Aimuratov",
    title = "{Quadrupolar gravitational fields described by the $q-$metric}",
    eprint = "1310.5339",
    archivePrefix = "arXiv",
    primaryClass = "gr-qc",
    journal = "Version published in International Journal of Mathematics and Physics",
    volume = "3",
    pages = "133",
    year = "2012"
}

@article{Fisher:1948yn,
    author = "Fisher, I. Z.",
    title = "{Scalar mesostatic field with regard for gravitational effects}",
    eprint = "gr-qc/9911008",
    archivePrefix = "arXiv",
    journal = "Zh. Eksp. Teor. Fiz.",
    volume = "18",
    pages = "636--640",
    year = "1948"
}

@article{Janis:1968zz,
    author = "Janis, Allen I. and Newman, Ezra T. and Winicour, Jeffrey",
    title = "{Reality of the Schwarzschild Singularity}",
    doi = "10.1103/PhysRevLett.20.878",
    journal = "Phys. Rev. Lett.",
    volume = "20",
    pages = "878--880",
    year = "1968"
}

@article{Gurtug:2021noy,
    author = "Gurtug, Ozay and Halilsoy, Mustafa and Mangut, Mert",
    title = "{The charged Zipoy{\textendash}Voorhees metric with astrophysical applications}",
    eprint = "2110.12188",
    archivePrefix = "arXiv",
    primaryClass = "gr-qc",
    doi = "10.1140/epjc/s10052-022-10626-0",
    journal = "Eur. Phys. J. C",
    volume = "82",
    number = "8",
    pages = "671",
    year = "2022"
}

@article{Majumdar:1947eu,
    author = "Majumdar, S. D.",
    title = "{A class of exact solutions of Einstein's field equations}",
    doi = "10.1103/PhysRev.72.390",
    journal = "Phys. Rev.",
    volume = "72",
    pages = "390--398",
    year = "1947"
}

@article{Papaetrou:1947ib,
    author = "Papaetrou, A.",
    title = "{A Static solution of the equations of the gravitational field for an arbitrary charge distribution}",
    journal = "Proc. Roy. Irish Acad. A",
    volume = "51",
    pages = "191--204",
    year = "1947"
}

@article{PhysRev.133.B845,
  title = {Electromagnetic Fields in a Homogeneous, Nonisotropic Universe},
  author = {Brill, Dieter R.},
  journal = {Phys. Rev.},
  volume = {133},
  issue = {3B},
  pages = {B845--B848},
  numpages = {0},
  year = {1964},
  month = {Feb},
  publisher = {American Physical Society},
  doi = {10.1103/PhysRev.133.B845},
  url = {https://link.aps.org/doi/10.1103/PhysRev.133.B845}
}

@article{Halilsoy:2022pzr,
    author = "Halilsoy, Mustafa and Mangut, Mert and Hsieh, Chia-Li",
    title = "{Stationary, charged Zipoy{\textendash}Voorhees metric from colliding wave spacetime}",
    eprint = "2210.15007",
    archivePrefix = "arXiv",
    primaryClass = "gr-qc",
    doi = "10.1007/s10714-023-03151-2",
    journal = "Gen. Rel. Grav.",
    volume = "55",
    number = "9",
    pages = "103",
    year = "2023"
}

@article{Yazadjiev:2000sa,
    author = "Yazadjiev, Stoytcho S.",
    editor = "Mateev, M.",
    title = "{Exact static solutions in Einstein-Maxwell dilaton gravity with arbitrary dilaton coupling parameter}",
    eprint = "gr-qc/0101078",
    archivePrefix = "arXiv",
    journal = "Bulg. J. Phys.",
    volume = "27N3",
    pages = "58--61",
    year = "2000"
}

@article{Porfyriadis:2023qqo,
    author = "Porfyriadis, Achilleas P. and Remmen, Grant N.",
    title = "{Charged dilatonic spacetimes in string theory}",
    eprint = "2301.08256",
    archivePrefix = "arXiv",
    primaryClass = "hep-th",
    reportNumber = "CCTP-2023-1, ITCP-IPP 2023/1",
    doi = "10.1007/JHEP03(2023)125",
    journal = "JHEP",
    volume = "03",
    pages = "125",
    year = "2023"
}

@article{Penney:1969xk,
    author = "Penney, R.",
    title = "{Generalization of the reissner-nordstroem solution to the einstein field equations}",
    doi = "10.1103/PhysRev.182.1383",
    journal = "Phys. Rev.",
    volume = "182",
    pages = "1383--1384",
    year = "1969"
}

@article{SJPoletti_1995,
doi = {10.1088/0264-9381/12/7/017},
url = {https://doi.org/10.1088/0264-9381/12/7/017},
year = {1995},
month = {jul},
publisher = {},
volume = {12},
number = {7},
pages = {1753},
author = {S J Poletti and J Twamley and D L Wiltshire},
title = {Dyonic dilaton black holes},
journal = {Classical and Quantum Gravity}
}

@article{Ellis:1973yv,
    author = "Ellis, H. G.",
    title = "{Ether flow through a drainhole - a particle model in general relativity}",
    doi = "10.1063/1.1666161",
    journal = "J. Math. Phys.",
    volume = "14",
    pages = "104--118",
    year = "1973"
}

@article{Bronnikov:1973fh,
    author = "Bronnikov, K. A.",
    title = "{Scalar-tensor theory and scalar charge}",
    journal = "Acta Phys. Polon. B",
    volume = "4",
    pages = "251--266",
    year = "1973"
}

@article{Gibbons:2016bok,
    author = "Gibbons, Gary W. and Volkov, Mikhail S.",
    title = "{Ring wormholes via duality rotations}",
    eprint = "1606.04879",
    archivePrefix = "arXiv",
    primaryClass = "hep-th",
    doi = "10.1016/j.physletb.2016.07.012",
    journal = "Phys. Lett. B",
    volume = "760",
    pages = "324--328",
    year = "2016"
}

@article{Volkov:2026rpu,
    author = "Volkov, Mikhail S.",
    title = "{Stationary generalizations for the vacuum ring wormhole}",
    eprint = "2605.27600",
    archivePrefix = "arXiv",
    primaryClass = "gr-qc",
    doi = "10.1103/mm69-kjkh",
    journal = "Phys. Rev. D",
    volume = "114",
    number = "8",
    pages = "084031",
    year = "2026"
}

@article{Sarmiento-Alvarado:2026efk,
  author = {Sarmiento-Alvarado, I. A. and Bixano, Leonel and Matos, Tonatiuh},
  title = {{Axially symmetric wormholes}},
  eprint = {2602.19975},
  archiveprefix = {arXiv},
  primaryclass = {gr-qc},
  doi = {10.1088/1361-6382/ae8a0b},
  journal = {Class. Quant. Grav.},
  volume = {43},
  number = {17},
  pages = {175004},
  year = {2026},
}

@article{Sarmiento-Alvarado:2026prq,
  author = {Sarmiento-Alvarado, I. A. and Bixano, Leonel and Matos, Tonatiuh},
  title = {{Exact Ricci-flat wormhole solution in five dimensions}},
  doi = {10.1088/1361-6382/ae8ef4},
  journal = {Class. Quant. Grav.},
  volume = {43},
  number = {15},
  pages = {155022},
  year = {2026},
}

\end{document}